\ifdefined\pdfoutput\pdfoutput=1\fi

\documentclass[11pt]{article}

\usepackage[preprint]{acl}

\usepackage{times}
\usepackage{latexsym}

\usepackage[T1]{fontenc}

\usepackage[utf8]{inputenc}

\usepackage{microtype}

\usepackage{inconsolata}

\usepackage{graphicx}

\usepackage{microtype}
\microtypecontext{spacing=nonfrench}

\usepackage{hyperref}
\usepackage{longtable}
\usepackage{arydshln}
\usepackage{bm}      
\usepackage{booktabs}
\usepackage{array}
\usepackage{makecell}
\usepackage{amsmath, amssymb, amsthm}
\usepackage{cleveref}  
\usepackage{amsmath, etoolbox}

\makeatletter
\patchcmd{\gathered@d@t@}
  {\tagform@{\thetag@form@}\fi}
  {%
    \setbox\z@\hbox{\tagform@{\thetag@form@}}%
    \dimen@=\ht\z@ \advance\dimen@ by \dp\z@
    \rlap{\raisebox{0.5\dimen@}{\tagform@{\thetag@form@}}}%
  }{}{}
\makeatother

\usepackage{algorithm}
\usepackage{algpseudocode}
\usepackage{ragged2e} 
\usepackage{graphicx}
\usepackage{xspace}
\usepackage{multirow}
\usepackage{pifont} 
\usepackage{colortbl} 
\PassOptionsToPackage{table}{xcolor} 
\usepackage{xcolor}
\usepackage{comment}
\usepackage{tabularx}

\newcommand{\detectorname}{\textsc{React}}
\newcommand{\modelname}{\textsc{Greater}}
\newcommand{\attackname}{\textsc{Greater-A}\xspace}
\newcommand{\defensename}{\textsc{Greater-D}\xspace}
\newcommand{\pecola}{\textsc{Pecola}}
\newcommand{\greencheckmark}{{\color{green}{\checkmark}}}
\newcommand{\redxmark}{{\color{red}{\ding{55}}}}
\newcommand{\zzh}[1]{{\color{olive} [ZH: #1]}}

\newcommand{\todo}[1]{\textbf{\color{red}{TODO: #1} }} 

\newcommand{\eg}{\emph{e.g.},\xspace}

\newcommand\figref[1]{Figure~\ref{#1}}

\newcommand\secref[1]{\S\ref{#1}}

\newcommand{\fakeparagraph}[1]
{\vspace{1mm}\noindent\textbf{#1}}
\definecolor{advbg}{RGB}{255,210,210} 

\title{\Large Fight Poison with Poison: Enhancing Robustness in Few-shot Machine-Generated Text Detection with Adversarial Training}

\author{Wenjing Duan\textsuperscript{$1,\dagger$}, Qi Zhou\textsuperscript{$1,\dagger$}, Yuanfan Li\textsuperscript{$1,\ast$} \\
        \textsuperscript{1}Faculty of Electronic and Information Engineering, Xi'an Jiaotong University\\ 
        \textsuperscript{$\dagger$} Equal contribution, \textsuperscript{$\ast$} Corresponding author\\
        \texttt{dwjduan@stu.xjtu.edu.cn}
        }

\begin{document}
\maketitle
\begin{abstract}
Machine-generated text (MGT) detection is critical for regulating online information ecosystems, yet existing detectors often underperform in few-shot settings and remain vulnerable to adversarial, humanizing attacks. 
To build accurate and robust detectors under limited supervision, we adopt a threat-modeling perspective and study detector vulnerabilities from an attacker’s viewpoint under an output-only black-box setting. 
Motivated by this perspective, we propose \textbf{R}AG-Guid\textbf{E}d \textbf{A}ttacker Strengthens \textbf{C}on\textbf{T}rastive Few-shot Detector (\detectorname), an adversarial training framework that improves both few-shot detection performance and robustness against attacks. 
\detectorname\ couples a humanization-oriented attacker with a target detector: the attacker leverages retrieval-augmented generation (RAG) to craft highly human-like adversarial examples to evade detection, while the detector learns from these adversaries with a contrastive objective to stabilize few-shot representation learning and enhance robustness. 
We alternately update the attacker and the detector to enable their co-evolution. 
Experiments on 4 datasets with 4 shot sizes and 3 random seeds show that \detectorname\ improves average detection F1 by \textbf{4.95} points over 8 state-of-the-art (SOTA) detectors and reduces the average attack success rate (ASR) under 4 strong attacks by \textbf{3.66} percentage points.
We will release our code and datasets upon acceptance.
\end{abstract}

\section{Introduction}
\label{sec:intro}
With the rapid advancement of large language models (LLMs) such as ChatGPT~\cite{achiam2023gpt}, LLaMA~\cite{dubey2024llama}, and DeepSeek~\cite{guo2025deepseek}, it has become straightforward to generate fluent, highly human-like text at scale. While this has greatly facilitated people’s daily lives, the misuse of generated content—such as LLM-written peer-review reports, phishing emails, and fake news—has raised widespread concern. To mitigate such abuse, researchers have developed a variety of methods for detecting machine-generated text (MGT)~\cite{liu2022coco, hu2023radar, liu2024does, bao2024fast, li2025iron} to inform readers about the provenance of a given passage.

\begin{figure*}[htbp]
	\centering
	\resizebox{1\textwidth}{!}{%
		\includegraphics{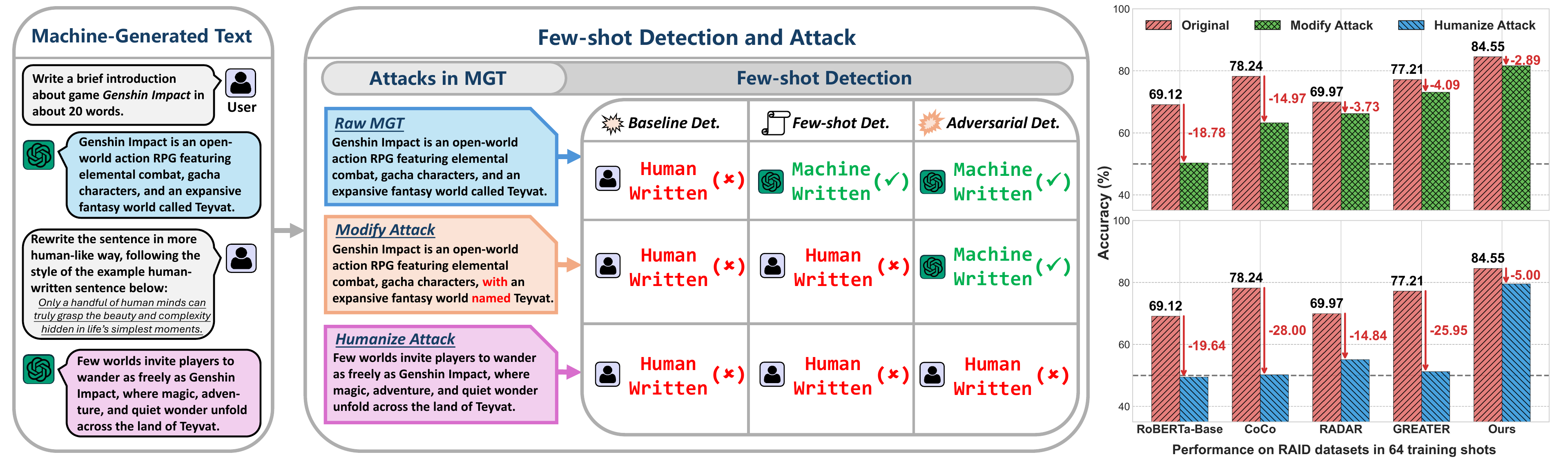} 
	}
	\caption{\textbf{Performance drop of model-based detectors under attacks in the few-shot settings.} The evaluated detectors include: (i) baseline detector (Baseline Det.), RoBERTa-Base~\cite{liu2019roberta}; (ii) few-shot-tailored detector (Few-shot Det.), CoCo~\cite{liu2022coco}; and (iii) adversarially trained detectors (Adversarial Det.), RADAR~\cite{hu2023radar}, \modelname~\cite{li2025iron} and \detectorname~(Ours). When trained on the RAID dataset~\cite{dugan2024raid} with 64 shots, the baseline detector and RADAR achieve only 60--70\% detection accuracy even on original texts, falling short of practical usability. Across prior detectors, performance under the \textit{Humanize} attack drops close to chance, whereas our \detectorname~degrades much less and remains substantially more robust.}
	\label{fig:fig1} 
	\vspace{-1.2em}
\end{figure*}

Modern MGT detectors can be broadly categorized into two types: (i) metric-based detectors and (ii) model-based detectors. 
Because metric-based detectors are often fragile to adversarial attacks~\cite{su2023detectllm,mitchell2023detectgpt,hans2024spotting}, model-based detectors are widely adopted in the literature and practice.
However, some recent studies~\cite{liu2022coco,liu2024does} show that training a high-performing MGT detector typically requires large-scale text data, which often limits its effectiveness in data-constrained settings.
As illustrated in Figure~\ref{fig:fig1}, in some few-shot settings, standard detectors typically achieve only 60--70\% detection accuracy, falling short of practical usability.
Few-shot-tailored detectors can partially improve performance, yet their accuracy still drops substantially even under simple attacks~\cite{wang2024stumbling}: minor edits to just a handful of words can cause misclassification.
Adversarial training is a common approach to improving robustness against attacks~\cite{hu2023radar,koike2024outfox,li2025iron}, but a more serious issue is that adversarially trained detectors can still degrade toward chance-level performance under stronger attacks (e.g., the \emph{Humanize} attack~\cite{wang2024stumbling}, which explicitly rewrites MGT to appear more human-like).
This reveals a key gap: robust few-shot MGT detection under strong attacks remains underexplored.

\noindent\textbf{Motivation.} We leverage \textit{adversarial training} to improve both detection accuracy and robustness of MGT detectors in few-shot settings.
However, existing adversarial training approaches~\cite{hu2023radar,koike2024outfox,li2025iron} often fail to remain robust against highly humanized attacks.
Since many attacks on MGT detectors revolve around \emph{humanizing} MGT to evade detection~\cite{krishna2023paraphrasing,koike2024outfox}, we are motivated by the intuition of \emph{``fight poison with poison''} and focus on constructing highly humanized adversarial examples to strengthen robustness under such attacks.
Nevertheless, na\"{i}vely incorporating adversarial examples can destabilize optimization in few-shot settings and hurt clean accuracy.
To address this, we draw on contrastive learning and develop a few-shot adversarial training framework that improves both robustness and detection performance.

\noindent \textbf{Our Work.} In this paper, we propose an adversarial training framework for training a robust MGT detector against attacks in few-shot settings, named 
\textbf{R}AG-Guid\textbf{E}d 
\textbf{A}ttacker Strengthens \textbf{C}on\textbf{T}rastive 
Few-shot Detector (\detectorname).
\detectorname~couples a detector that discriminates human-written texts (HWTs) from MGTs with an attacker that generates humanized adversarial examples to evade detection. 
To strengthen the attacker, we adopt a retrieval-augmented generation (RAG)~\cite{lewis2020retrieval} scheme: a retrieval pool built from training samples and previously generated adversaries provides a positive/negative pair as in-context demonstrations, encouraging more human-like perturbations. Under the black-box constraint where only detector \textbf{outputs} are available, we use an open-source surrogate model to approximate the target detector and to derive retrieval labels. To stabilize and accelerate few-shot learning, we further introduce a contrastive objective for the detector.
At each training step, we alternately update the detector and the attacker's surrogate model, enabling the attacker to generate increasingly humanized adversarial examples over time. Experimental results on 4 datasets with 3 random seeds and 4 shot settings demonstrate that \detectorname~achieves an average detection F1 score of \textbf{93.72}, outperforming SOTA detectors by \textbf{4.95} points. Moreover, \detectorname~attains an average attack success rate (ASR) of \textbf{10.04\%} under 4 attacks, reducing ASR by \textbf{3.66\%} compared to SOTA baselines. Our contributions are as follows: 

\begin{itemize}
    \item \textbf{Effective Adversarial Training Framework.} We propose an adversarial training framework, \detectorname, to improve the robustness of MGT detectors against attacks in few-shot settings. In this framework, the attacker produces highly humanized adversarial examples, while the detector employs contrastive learning over original and adversarial samples to accurately distinguish HWTs from MGTs and remain robust under attack.
    
    \item \textbf{Adversarial Example Generation.} We propose a novel adversarial example generation strategy in the black-box setting. We employ RAG to produce highly human-like texts, and use a surrogate model to obtain stronger positive--negative exemplar pairs for retrieval, further improving the quality of RAG-guided humanization.
    
    \item \textbf{Outstanding Performance.} Experimental results across 4 datasets, 3 random seeds, and 4 shot budgets demonstrate that our method surpasses 8 SOTA detectors in detection accuracy and exhibits stronger robustness against 4 powerful attacks.
    
\end{itemize}

\section{Related Work}

\noindent \textbf{Machine-Generated Text (MGT) Detectors.} 
With the rapid development of LLMs, numerous detectors have been proposed to identify machine-generated text (MGT). Existing approaches broadly fall into two categories: \emph{metric-based} and \emph{model-based} detectors. Metric-based detectors~\cite{su2023detectllm,bao2024fast,hans2024spotting,xu2024training,zhu2025dna} require little or no supervised training and classify text using statistical signals of MGT, but they can be fragile to adversarial perturbations and often produce uncalibrated scores, complicating threshold selection in deployment. In contrast, model-based detectors~\cite{liu2022coco,hu2023radar,liu2024does,koike2024outfox,li2025iron} fine-tune a text classifier (e.g., BERT/RoBERTa) and are generally stronger, yet they usually require substantial labeled data to achieve high accuracy. Few-shot-tailored detectors~\cite{liu2022coco,liu2024does} mitigate the data requirement but remain vulnerable to attacks, while adversarially trained detectors~\cite{hu2023radar,li2025iron} may suffer optimization instability and reduced clean accuracy in few-shot settings. To address these limitations, our work focuses on improving both detection accuracy and robustness against attacks under limited supervision.

\noindent \textbf{Adversarial Training.} Adversarial training optimizes a detector to preserve correct predictions on adversarial examples that are deliberately crafted to induce misclassification. However, recent studies~\cite{thorat2025dactyl,umrajkar2025dac} suggest that existing adversarial training methods do not transfer well to few-shot detection: directly incorporating adversarial examples can exacerbate underfitting and degrade accuracy when supervision is scarce. Moreover, adversarial training for MGT detectors~\cite{hu2023radar,koike2024outfox,li2025iron} remains ineffective against highly humanized, strong attacks in few-shot settings. To address these limitations, our method leverages RAG~\cite{lewis2020retrieval} to generate highly human-like adversarial examples and introduces a contrastive objective to stabilize few-shot representation learning, improving both accuracy and robustness.

\section{Threat Model}
We follow the standard threat modeling framework outlined in prior work~\cite{biggio2018wild,hu2023radar,koike2024outfox,li2025iron} and describe the goals and capabilities of both the attacker and the detector in our adversarial-training setup.

\noindent\textbf{Attacker’s Goals and Capabilities.}
\label{goal}
Given an MGT passage, the attacker aims to \emph{humanize} it to deceive the detector and outputs a rewrite.
We call rewrites \emph{from MGT inputs} as \emph{adversarial examples}, regardless of success.
During training only, we also generate auxiliary rewrites for HWTs as data augmentation and treat them as \emph{MGT-labeled} samples for detector optimization.
We do not regard these HWT-based rewrites as attacks, and they never define the attacker’s objective.
To reflect real-world deployment where the detector is accessed only via an API that returns \textbf{top-1 predicted labels}, we consider a black-box setting in which the attacker has no access to the detector’s weights, architecture, or any proprietary training data beyond the provided few-shot set, and can observe only the output labels via queries.
The attacker is allowed to leverage (i) any publicly available LMs/LLMs, including both open-source models and closed-source/commercial models accessed via APIs, and (ii) previously generated rewrites together with detector-returned labels to synthesize stronger adversarial examples.
The attacker makes \textbf{one} detector query per generated rewrite (including auxiliary rewrites during training).

\noindent\textbf{Detector’s Goals and Capabilities.}
The detector aims to correctly distinguish HWTs from MGTs while remaining robust to adversarial examples.
During training, the detector has access only to a few-shot labeled training set (containing both HWTs and MGTs), the adversarial examples produced from MGT inputs, and the auxiliary rewrites generated from HWT inputs for data augmentation.
We further assume that the detector has no \emph{direct access} to LLMs or large-scale external corpora beyond the provided training data.
We emphasize that any LLM usage is confined to the attacker module for generating rewrites, while the deployed detector itself never queries LLMs and only optimizes on the resulting texts.

\section{Methodology} 
\label{sec:methodology}

\begin{figure*}[htbp]
	\centering
	\resizebox{1\textwidth}{!}{%
		\includegraphics{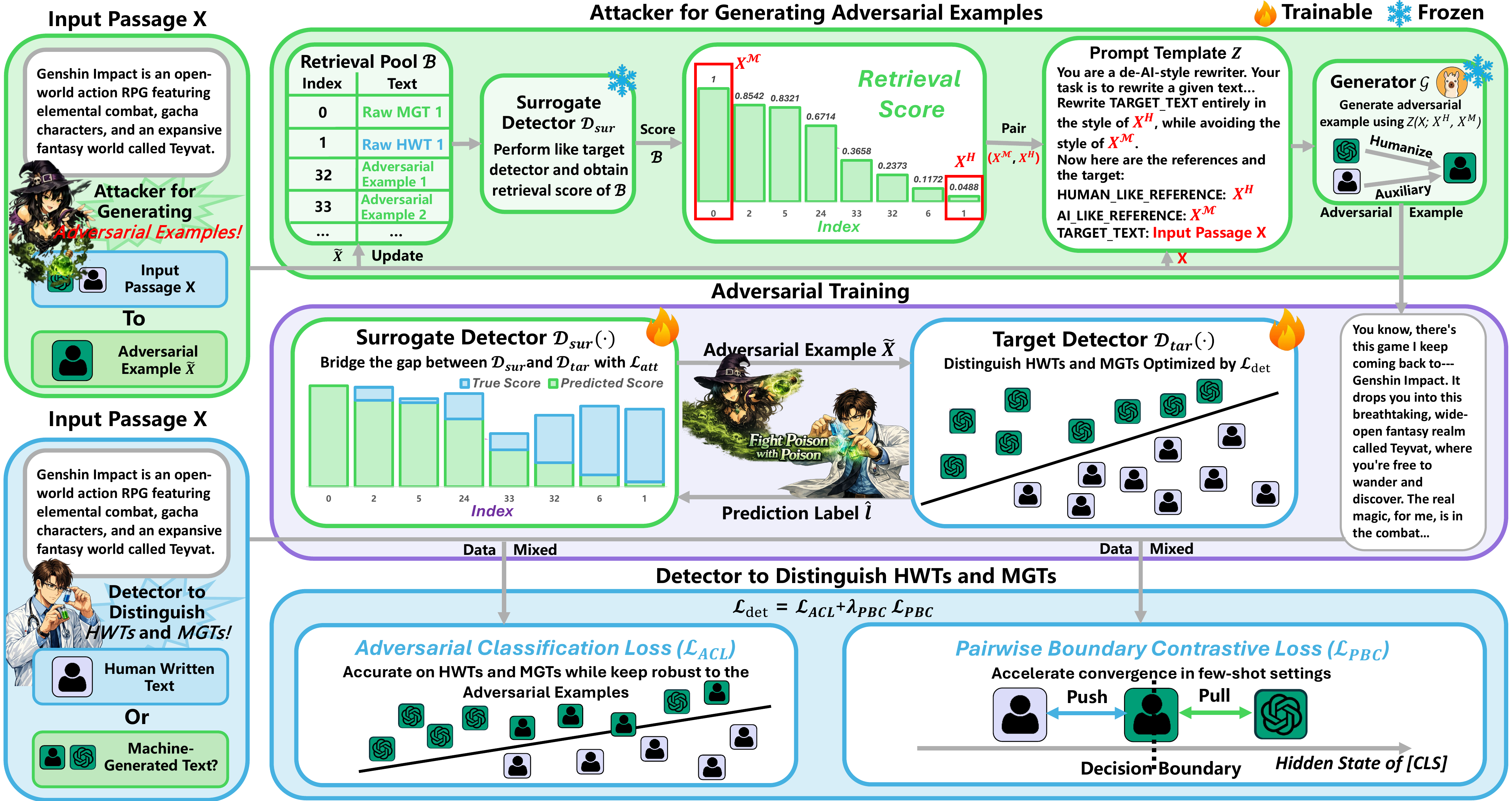} 
	}
	\caption{\textbf{Pipeline of \detectorname.}
		The attacker generates adversarial examples via RAG (\secref{sec:attacker}), which are then fed into the detector (\secref{sec:detector}) and used in the adversarial training procedure to update the retrieval pool. (\secref{AT}).}
	\label{fig:framework} 
	\vspace{-1.2em}
\end{figure*}

In this section, we introduce our \detectorname~framework. The architecture of \detectorname~is shown in Figure~\ref{fig:framework}. Our \detectorname~contains an attacker and a detector. In the following subsections, we first describe the workflows of our attacker and detector, and then  introduce the adversarial training procedure.

\subsection{Attacker for Generating Adversarial Examples}
\label{sec:attacker}

Our attacker consists of a surrogate detector $\mathcal{D}_{\mathrm{sur}}(\cdot)$ and a generator $\mathcal{G}(\cdot)$.
To achieve the attacker's goal in \secref{goal}, we adopt a RAG-style humanization procedure to generate adversarial candidates, and update $\mathcal{D}_{\mathrm{sur}}$ with an attacker loss after each training step to keep the retrieval scoring aligned with the target detector.

\noindent \textbf{Retrieval-Augmented Generation.}
\label{sec:rag}
Given an input passage $X$, prior work often crafts an adversarial example $\tilde{X}$\footnote{We use $\tilde{X}$ to denote a \emph{rewrite} generated by the attacker for an input passage $X$.
When $X$ is an MGT, $\tilde{X}$ is an \emph{adversarial example}; when $X$ is an HWT (during training only), $\tilde{X}$ is an \emph{auxiliary rewrite} used for data augmentation and is not considered an attack.} via local perturbations such as synonym substitution~\cite{jin2020bert,zhou2024humanizing} or mask-based replacement~\cite{li2020bert}, which may introduce unnatural phrasing and reduce readability.
Instead, we generate $\tilde{X}$ by conditioning $\mathcal{G}$ on retrieved exemplars.

Specifically, we maintain a text-only retrieval pool (knowledge base)
$\mathcal{B}=\{X_i\}_{i=1}^{|\mathcal{B}|}$,
initialized with the few-shot training passages and continuously expanded with attacker-generated rewrites from previous steps.
Let \(
P_{\mathrm{sur}}\bigl(l \mid X_i\bigr)
\)
denote the probability assigned to label $l$ by the surrogate, for retrieval, we score each $X_i$ with $\mathcal{D}_{\mathrm{sur}}$ by its MGT probability
$p_i = P_{\mathrm{sur}}(l=\mathrm{MGT}\mid X_i)$,
and use it as a scalar retrieval score $\mathbf{v}(X_i)=p_i$ to rank candidates in $\mathcal{B}$:
\begin{equation}
	\label{eq:eq1}
	\begin{split}
		X^{H}=\arg\min_{X_i\in\mathcal{B}} \mathbf{v}(X_i),\\
		X^{M}=\arg\max_{X_i\in\mathcal{B}} \mathbf{v}(X_i).
	\end{split}
\end{equation}
Here, $X^{H}$ and $X^{M}$ correspond to the most human-like and most machine-like instances under $\mathcal{D}_{\mathrm{sur}}$, serving as style anchors for rewriting.
This retrieved pair provides in-context exemplars for RAG: $X^{H}$ supplies human-writing cues to imitate, while $X^{M}$ highlights machine-like patterns to avoid.
We find that such \emph{extreme} anchors provide the clearest style contrast under the surrogate's scoring and we further study alternative prompting strategies in~\secref{sec:generation}.
We then inject $(X^{H},X^{M})$ into a prompt template $\mathcal{Z}(\cdot)$ and feed it to $\mathcal{G}$ to produce:
\begin{equation}
	\label{eq:eq2}
	\tilde{X}=\mathcal{G}\!\left(\mathcal{Z}(X;X^{H},X^{M})\right).
\end{equation}

For future retrieval, we insert only the rewritten text into the pool, i.e., $\mathcal{B}\leftarrow \mathcal{B}\cup\{\tilde{X}\}$, yielding a self-improving retrieval pool.
We demonstrate in Appendix~\ref{sec:human} via human evaluation that our RAG strategy can effectively generate humanized texts, and we further provide a case study in Appendix~\ref{sec:case}.

\noindent\textbf{Attacker's Loss.}
To align $\mathcal{D}_{\mathrm{sur}}$ with the label-only feedback from $\mathcal{D}_{\mathrm{tar}}$ in the black-box setting, for each generated rewrite $\tilde{X}$ we take
$\hat{l}=\arg\max \mathcal{D}_{\mathrm{tar}}(\tilde{X})$
as a pseudo label and minimize the negative log-likelihood under $\mathcal{D}_{\mathrm{sur}}$:
\begin{equation}
	\label{eq:att}
	\mathcal{L}_{\text{att}}
	=
	-\log P_{\mathrm{sur}}\bigl(\hat{l} \mid \tilde{X}\bigr).
\end{equation}
In each training step, we update the surrogate model using Eq.~\eqref{eq:att} while keeping the detector frozen. Detailed procedures are described in~\secref{AT}.

\subsection{Detector to Distinguish HWTs and MGTs}
\label{sec:detector}

Our detector is a target classifier $\mathcal{D}_{\mathrm{tar}}(\cdot)$.
Following the detector goal in \secref{goal}, we optimize $\mathcal{D}_{\mathrm{tar}}$ with two complementary objectives: an adversarial classification loss that enforces label correctness under humanization, and a pairwise boundary contrastive loss that regularizes the representation geometry.

\noindent\textbf{Adversarial Classification Loss.}
The Adversarial Classification Loss (ACL) trains $\mathcal{D}_{\mathrm{tar}}$ to correctly classify the original training sample while remaining robust to the attacker-produced humanized rewrite.
Given $(X,l)\in D_{\mathrm{train}}$ and its attacker-generated rewrite $\tilde{X}$, we treat $\tilde{X}$ as an \emph{MGT-labeled} instance for training (both for adversarial examples from MGT inputs and auxiliary rewrites from HWT inputs).
Let $P_{\mathrm{tar}}(l\mid X)$ be the probability assigned by $\mathcal{D}_{\mathrm{tar}}$ to label $l$ for input $X$.
We define
\begin{equation}
	\small
	\label{eq:acl}
	\mathcal{L}_{\mathrm{ACL}}
	=
	-\log P_{\mathrm{tar}}(l \mid X)
	-\alpha \log P_{\mathrm{tar}}(\mathrm{MGT} \mid \tilde{X}),
\end{equation}
where $\alpha$ down-weights the adversarial term.
We introduce $\alpha$ because the attacker continuously generates additional MGT-labeled rewrites, which effectively amplifies class imbalance and can destabilize optimization and bias training toward adversarial artifacts.
By controlling the contribution of $\tilde{X}$, $\alpha$ stabilizes optimization and improves robustness transfer to newly generated adversarial candidates.

\noindent\textbf{Pairwise Boundary Contrastive Loss.}
To further stabilize optimization and accelerate convergence in few-shot settings, we propose a Pairwise Boundary Contrastive loss (PBC) that directly constrains the geometry between an input and its rewrite.
While prior work often employs InfoNCE~\cite{liu2024does} or supervised contrastive learning~\cite{liu2022coco}, these objectives typically rely on many informative in-batch negatives or a memory bank to construct reliable contrast sets.
Such requirements become brittle under few-shot data and small effective batch sizes, introducing additional components whose gains are sensitive to design choices.
We instead use a pairwise formulation on a single $(X,\tilde{X})$ pair, which is more sample-efficient and less dependent on batch composition.

\begin{algorithm}[t]
	\small
	\caption{Adversarial Training Procedure}
	\label{alg:adv_train}
	\begin{algorithmic}[1]
		\State \textbf{Input:} few-shot training set $D_{\mathrm{train}}=\{(X,l)\}$,
		surrogate detector $\mathcal{D}_{\mathrm{sur}}$,
		target detector $\mathcal{D}_{\mathrm{tar}}$,
		generator $\mathcal{G}$,
		prompt template $\mathcal{Z}(\cdot)$,
		maximum epochs $epoch_{\max}$,
		hyperparameters $\alpha$, $\lambda_{\mathrm{pbc}}$, $\delta_{\mathrm{same}}$, $\delta_{\mathrm{diff}}$
		\State \textbf{Initialize:} retrieval pool $\mathcal{B}\leftarrow\{X \mid (X,l)\in D_{\mathrm{train}}\}$, $epoch \leftarrow 0$
		\begin{center}
			\textit{*** adversarial training phase begins ***}
		\end{center}
		\While{$epoch < epoch_{\max}$}
		\For{each training pair $(X,l)$ in $D_{\mathrm{train}}$} 
		\begin{center}
			\textit{*** Step A: RAG-based humanization. ***}
		\end{center}
		\State Compute retrieval scores for all $X_i\in\mathcal{B}$ using $\mathcal{D}_{\mathrm{sur}}$
		\State Retrieve style anchors $(X^{H},X^{M})$ by Eq.~\eqref{eq:eq1}
		\State Build the prompt $\mathcal{Z}(X;X^{H},X^{M})$
		\State Generate the rewrite $\tilde{X}$ by Eq.~\eqref{eq:eq2}
		\State Query $\mathcal{D}_{\mathrm{tar}}$ once, obtain $\hat{l}\leftarrow \arg\max \mathcal{D}_{\mathrm{tar}}(\tilde{X})$
		\begin{center}
			\textit{*** Step B: Update $\mathcal{D}_{\mathrm{sur}}$***}
		\end{center}		
		\State Calculate attacker loss $\mathcal{L}_{\mathrm{att}}$ using Eq.~\eqref{eq:att}
		\State Update $\mathcal{D}_{\mathrm{sur}}$ with SGD ($\mathcal{D}_{\mathrm{tar}}$ frozen)
		\begin{center}
			\textit{*** Step C: Update $\mathcal{D}_{\mathrm{tar}}$***}
		\end{center}	
		\State Assign the training label for the rewrite $\tilde{X}$ as $\tilde{l}\leftarrow \mathrm{MGT}$.
		\State Calculate $\mathcal{L}_{\mathrm{ACL}}$ using Eq.~\eqref{eq:acl}
		\State Calculate $\mathcal{L}_{\mathrm{PBC}}$ using Eq.~\eqref{eq:pbc}
		\State  $\mathcal{L}_{\mathrm{det}}\leftarrow \mathcal{L}_{\mathrm{ACL}}+\lambda_{\mathrm{pbc}}\mathcal{L}_{\mathrm{PBC}}$ (Eq.~\eqref{eq:det_total}) 
		\State Update $\mathcal{D}_{\mathrm{tar}}$ with SGD
		($\mathcal{D}_{\mathrm{sur}}$ frozen)
		\begin{center}
			\textit{*** Step D: Update $\mathcal{B}$***}
		\end{center}	
		\State Update the pool $\mathcal{B}\leftarrow \mathcal{B}\cup\{\tilde{X}\}$
		\EndFor
		\State $epoch \leftarrow epoch + 1$
		\EndWhile
		\State \textbf{Output:} trained detector $\mathcal{D}_{\mathrm{tar}}$ and the updated pool $\mathcal{B}$
	\end{algorithmic}
\end{algorithm}
Let $\mathbf{h}$ and $\tilde{\mathbf{h}}$ be the last-layer \texttt{[CLS]} representations of $X$ and $\tilde{X}$ from $\mathcal{D}_{\mathrm{tar}}$, and let
$c=\cos(\mathrm{norm}(\mathbf{h}),\mathrm{norm}(\tilde{\mathbf{h}}))$ be their cosine similarity, where $\mathrm{norm}(\mathbf{u})=\mathbf{u}/\lVert\mathbf{u}\rVert_2$ denotes $\ell_2$ normalization.
We define
\begin{equation}
	\label{eq:pbc}
	\begin{aligned}
		\mathcal{L}_{\mathrm{PBC}}
		&=
		\mathbb{I}[l=\tilde{l}]\cdot \max\!\bigl(0,(1-\delta_{\mathrm{same}})-c\bigr) \\
		&\quad+
		\mathbb{I}[l\neq \tilde{l}]\cdot \max\!\bigl(0,c-\delta_{\mathrm{diff}}\bigr),
	\end{aligned}
\end{equation}
where $\mathbb{I}[\cdot]$ is the indicator function that equals $1$ if the condition holds and $0$ otherwise, $l$ and $\tilde{l}$ denote the labels of $X$ and $\tilde{X}$, and $\delta_{\mathrm{same}}$ and $\delta_{\mathrm{diff}}$ are the margins for same-label and different-label pairs.
Minimizing Eq.~\eqref{eq:pbc} enforces $c \ge 1-\delta_{\mathrm{same}}$ for same-label pairs and $c \le \delta_{\mathrm{diff}}$ for different-label pairs, yielding more stable decision boundaries.
This low-variance, pairwise signal is particularly effective for few-shot learning, improving convergence without requiring large negative sets.

\noindent\textbf{Total Loss.}
The overall detector objective is
\begin{equation}
	\label{eq:det_total}
	\mathcal{L}_{\mathrm{det}}
	=
	\mathcal{L}_{\mathrm{ACL}}
	+
	\lambda_{\mathrm{pbc}}\mathcal{L}_{\mathrm{PBC}},
\end{equation}
where $\lambda_{\mathrm{pbc}}$ weights the contrastive term.
At each training step, we update $\mathcal{D}_{\mathrm{tar}}$ by minimizing Eq.~\eqref{eq:det_total} with the attacker frozen, as detailed in \secref{AT}.

\begin{table*}[t]
	\centering
	\renewcommand\arraystretch{1.2}
	\footnotesize
	\resizebox{\textwidth}{!}{%
		\begin{tabular}{c c c c c c c c c c c c}
			\toprule
			\multirow{2}{*}{\textbf{Dataset}} & \multirow{2}{*}{\textbf{Shot}} & \multirow{2}{*}{\textbf{Metric}}
			& \multicolumn{3}{c}{\textbf{Metric-based Detectors}}
			& \multicolumn{6}{c}{\textbf{Model-based Detectors}} \\
			\cmidrule(lr){4-6}\cmidrule(lr){7-12}
			& & & \textbf{Binoculars} & \textbf{Fast DetectGPT} & \textbf{LRR}
			& \textbf{RoBERTa-Base} & \textbf{CoCo$^{\dagger}$} & \textbf{\pecola$^{\dagger}$} & \textbf{RADAR*} & \textbf{\modelname*} & \textbf{\detectorname~(Ours)*} \\ 
			\midrule
			\multirow{10}{*}{\textbf{DetectRL}} & \multirow{2}{*}{\textit{32}} & \textit{Acc} $\uparrow$ & 88.38$_{\pm 0.06}$ & 86.90$_{\pm 0.22}$ & 76.47$_{\pm 0.68}$ & 86.08$_{\pm 0.96}$ & \underline{92.20$_{\pm 3.42}$} & 50.24$_{\pm 0.42}$ & 87.76$_{\pm 12.49}$ & 90.17$_{\pm 1.64}$ & \textbf{99.46}$_{\bm{\pm 0.58}}$ \\
			&  & \textit{F1} $\uparrow$ & 87.62$_{\pm 0.61}$ & 86.10$_{\pm 0.59}$ & 73.04$_{\pm 2.60}$ & 83.82$_{\pm 1.30}$ & \underline{91.48$_{\pm 4.08}$} & 0.96$_{\pm 1.67}$ & 89.00$_{\pm 9.98}$ & 89.11$_{\pm 2.02}$ & \textbf{99.46}$_{\bm{\pm 0.58}}$ \\
			\cdashline{2-12}[1pt/1pt]
			& \multirow{2}{*}{\textit{64}} & \textit{Acc} $\uparrow$ & 88.50$_{\pm 0.49}$ & 86.65$_{\pm 0.52}$ & 76.93$_{\pm 0.58}$ & \underline{99.43$_{\pm 0.61}$} & 89.83$_{\pm 1.38}$ & 93.78$_{\pm 5.48}$ & 99.06$_{\pm 0.71}$ & 99.12$_{\pm 0.27}$ & \textbf{99.82}$_{\bm{\pm 0.27}}$ \\
			&  & \textit{F1} $\uparrow$ & 88.01$_{\pm 0.57}$ & 86.02$_{\pm 0.05}$ & 74.37$_{\pm 1.68}$ & \underline{99.43$_{\pm 0.62}$} & 88.66$_{\pm 1.71}$ & 93.16$_{\pm 6.50}$ & 99.06$_{\pm 0.69}$ & 99.11$_{\pm 0.28}$ & \textbf{99.82}$_{\bm{\pm 0.27}}$ \\
			\cdashline{2-12}[1pt/1pt]
			& \multirow{2}{*}{\textit{128}} & \textit{Acc} $\uparrow$ & 88.60$_{\pm 0.52}$ & 86.97$_{\pm 0.06}$ & 77.10$_{\pm 0.39}$ & 95.54$_{\pm 6.69}$ & 99.61$_{\pm 0.15}$ & \textbf{99.93}$_{\bm{\pm 0.11}}$ & 98.70$_{\pm 1.65}$ & 99.48$_{\pm 0.27}$ & \underline{99.74$_{\pm 0.07}$} \\
			&  & \textit{F1} $\uparrow$ & 88.14$_{\pm 0.17}$ & 86.17$_{\pm 0.30}$ & 74.69$_{\pm 1.30}$ & 94.98$_{\pm 7.66}$ & 99.61$_{\pm 0.15}$ & \textbf{99.93}$_{\bm{\pm 0.11}}$ & 98.66$_{\pm 1.71}$ & 99.48$_{\pm 0.27}$ & \underline{99.74$_{\pm 0.08}$} \\
			\cdashline{2-12}[1pt/1pt]
			& \multirow{2}{*}{\textit{256}} & \textit{Acc} $\uparrow$ & 88.87$_{\pm 0.18}$ & 87.03$_{\pm 0.13}$ & 77.40$_{\pm 0.13}$ & 98.08$_{\pm 2.78}$ & 99.87$_{\pm 0.07}$ & \textbf{99.97}$_{\bm{\pm 0.03}}$ & 98.94$_{\pm 1.31}$ & \underline{99.95$_{\pm 0.05}$} & 99.79$_{\pm 0.16}$ \\
			&  & \textit{F1} $\uparrow$ & 88.38$_{\pm 0.22}$ & 86.23$_{\pm 0.44}$ & 75.02$_{\pm 0.37}$ & 97.99$_{\pm 2.94}$ & 99.87$_{\pm 0.07}$ & \textbf{99.97}$_{\bm{\pm 0.03}}$ & 98.92$_{\pm 1.35}$ & \underline{99.95$_{\pm 0.05}$} & 99.79$_{\pm 0.16}$ \\
			\cdashline{2-12}[1pt/1pt]
			& \multirow{2}{*}{\textbf{\textit{Avg.}}} & \textit{Acc} $\uparrow$ & 88.59$_{\pm 0.34}$ & 86.84$_{\pm 0.23}$ & 76.97$_{\pm 0.44}$ & 94.78$_{\pm 2.76}$ & 95.38$_{\pm 1.26}$ & 85.98$_{\pm 1.51}$ & 96.14$_{\pm 4.04}$ & \underline{97.15$_{\pm 0.56}$} & \textbf{99.70}$_{\bm{\pm 0.27}}$ \\
			&  & \textit{F1} $\uparrow$ & 88.04$_{\pm 0.39}$ & 86.13$_{\pm 0.32}$ & 74.28$_{\pm 1.46}$ & 94.05$_{\pm 3.13}$ & 94.91$_{\pm 1.52}$ & 73.50$_{\pm 2.00}$ & 96.41$_{\pm 3.43}$ & \underline{96.91$_{\pm 0.68}$} & \textbf{99.70}$_{\bm{\pm 0.27}}$ \\
			
			\midrule
			
			\multirow{10}{*}{\textbf{OUTFOX}} & \multirow{2}{*}{\textit{32}} & \textit{Acc} $\uparrow$ & 86.62$_{\pm 0.51}$ & 80.45$_{\pm 0.28}$ & 78.52$_{\pm 0.29}$ & \underline{88.15$_{\pm 2.78}$} & 53.17$_{\pm 2.69}$ & 56.07$_{\pm 10.47}$ & 77.34$_{\pm 16.01}$ & 68.85$_{\pm 9.14}$ & \textbf{95.64}$_{\bm{\pm 1.56}}$ \\
			&  & \textit{F1} $\uparrow$ & 85.96$_{\pm 0.74}$ & 78.43$_{\pm 0.28}$ & 77.14$_{\pm 1.12}$ & \underline{88.48$_{\pm 2.00}$} & 68.12$_{\pm 1.24}$ & 23.16$_{\pm 39.94}$ & 82.51$_{\pm 11.60}$ & 76.20$_{\pm 5.09}$ & \textbf{95.77}$_{\bm{\pm 1.42}}$ \\
			\cdashline{2-12}[1pt/1pt]
			& \multirow{2}{*}{\textit{64}} & \textit{Acc} $\uparrow$ & 86.67$_{\pm 0.45}$ & 80.38$_{\pm 0.19}$ & 78.42$_{\pm 0.33}$ & 90.84$_{\pm 1.74}$ & 93.31$_{\pm 1.23}$ & 67.69$_{\pm 15.73}$ & \underline{93.62$_{\pm 3.18}$} & 88.10$_{\pm 4.06}$ & \textbf{97.12}$_{\bm{\pm 0.05}}$ \\
			&  & \textit{F1} $\uparrow$ & 86.07$_{\pm 0.79}$ & 78.19$_{\pm 0.10}$ & 77.16$_{\pm 1.46}$ & 90.54$_{\pm 2.29}$ & \underline{93.40$_{\pm 0.90}$} & 67.88$_{\pm 8.74}$ & 93.33$_{\pm 3.81}$ & 88.04$_{\pm 3.66}$ & \textbf{97.16}$_{\bm{\pm 0.02}}$ \\
			\cdashline{2-12}[1pt/1pt]
			& \multirow{2}{*}{\textit{128}} & \textit{Acc} $\uparrow$ & 86.70$_{\pm 0.61}$ & 80.20$_{\pm 0.05}$ & 78.53$_{\pm 0.25}$ & 87.65$_{\pm 9.64}$ & \underline{95.80$_{\pm 1.94}$} & 92.76$_{\pm 5.57}$ & 95.30$_{\pm 2.26}$ & 95.57$_{\pm 2.65}$ & \textbf{96.73}$_{\bm{\pm 0.48}}$ \\
			&  & \textit{F1} $\uparrow$ & 86.06$_{\pm 0.85}$ & 78.18$_{\pm 0.09}$ & 77.02$_{\pm 0.85}$ & 88.01$_{\pm 8.84}$ & \underline{95.64$_{\pm 2.12}$} & 93.29$_{\pm 4.62}$ & 95.39$_{\pm 2.10}$ & 95.38$_{\pm 2.93}$ & \textbf{96.75}$_{\bm{\pm 0.56}}$ \\
			\cdashline{2-12}[1pt/1pt]
			& \multirow{2}{*}{\textit{256}} & \textit{Acc} $\uparrow$ & 86.82$_{\pm 0.13}$ & 80.27$_{\pm 0.03}$ & 78.35$_{\pm 0.30}$ & 97.56$_{\pm 0.82}$ & 97.93$_{\pm 0.37}$ & \textbf{98.14}$_{\bm{\pm 0.55}}$ & 97.14$_{\pm 0.92}$ & 98.01$_{\pm 0.73}$ & \underline{98.03$_{\pm 0.52}$} \\
			&  & \textit{F1} $\uparrow$ & 86.21$_{\pm 0.15}$ & 78.22$_{\pm 0.10}$ & 76.69$_{\pm 1.03}$ & 97.55$_{\pm 0.81}$ & 97.93$_{\pm 0.38}$ & \textbf{98.15}$_{\bm{\pm 0.55}}$ & 97.10$_{\pm 0.98}$ & 98.02$_{\pm 0.73}$ & \underline{98.04}$_{\pm 0.51}$ \\
			\cdashline{2-12}[1pt/1pt]
			& \multirow{2}{*}{\textbf{\textit{Avg.}}} & \textit{Acc} $\uparrow$ & 86.70$_{\pm 0.42}$ & 80.33$_{\pm 0.14}$ & 78.45$_{\pm 0.29}$ & \underline{91.05$_{\pm 3.74}$} & 85.05$_{\pm 1.55}$ & 78.62$_{\pm 8.08}$ & 90.85$_{\pm 5.61}$ & 87.68$_{\pm 4.15}$ & \textbf{96.88}$_{\bm{\pm 0.65}}$ \\
			&  & \textit{F1} $\uparrow$ & 86.08$_{\pm 0.63}$ & 78.26$_{\pm 0.14}$ & 77.00$_{\pm 1.11}$ & 91.14$_{\pm 3.49}$ & 88.77$_{\pm 1.16}$ & 70.62$_{\pm 13.46}$ & \underline{92.08$_{\pm 4.60}$} & 89.41$_{\pm 3.14}$ & \textbf{96.94}$_{\bm{\pm 0.62}}$ \\
			
			\midrule
			
			\multirow{10}{*}{\textbf{RAID}} & \multirow{2}{*}{\textit{32}} & \textit{Acc} $\uparrow$ & \underline{84.07$_{\pm 0.16}$} & \textbf{86.40}$_{\bm{\pm 0.48}}$ & 74.19$_{\pm 0.61}$ & 72.09$_{\pm 4.59}$ & 53.40$_{\pm 2.92}$ & 50.73$_{\pm 1.06}$ & 56.85$_{\pm 10.44}$ & 53.91$_{\pm 5.76}$ & 79.30$_{\pm 0.90}$ \\
			&  & \textit{F1} $\uparrow$ & \underline{82.31$_{\pm 0.20}$} & \textbf{86.25}$_{\bm{\pm 0.32}}$ & 70.36$_{\pm 1.91}$ & 65.97$_{\pm 13.25}$ & 66.99$_{\pm 1.25}$ & 10.31$_{\pm 16.39}$ & 65.78$_{\pm 2.02}$ & 65.94$_{\pm 1.39}$ & 80.18$_{\pm 0.86}$ \\
			\cdashline{2-12}[1pt/1pt]
			& \multirow{2}{*}{\textit{64}} & \textit{Acc} $\uparrow$ & 83.37$_{\pm 1.37}$ & \textbf{86.42}$_{\bm{\pm 0.38}}$ & 74.46$_{\pm 0.42}$ & 69.12$_{\pm 6.59}$ & 78.24$_{\pm 4.63}$ & 50.00$_{\pm 0.00}$ & 69.97$_{\pm 17.30}$ & 77.21$_{\pm 12.42}$ & \underline{84.55$_{\pm 3.78}$} \\
			&  & \textit{F1} $\uparrow$ & 81.80$_{\pm 0.75}$ & \textbf{86.23}$_{\bm{\pm 0.25}}$ & 71.27$_{\pm 1.47}$ & 61.36$_{\pm 17.70}$ & 78.22$_{\pm 5.30}$ & 44.44$_{\pm 38.49}$ & 74.05$_{\pm 6.48}$ & 80.12$_{\pm 6.46}$ & \underline{83.63$_{\pm 4.11}$} \\
			\cdashline{2-12}[1pt/1pt]
			& \multirow{2}{*}{\textit{128}} & \textit{Acc} $\uparrow$ & 83.56$_{\pm 0.91}$ & 86.45$_{\pm 0.35}$ & 74.80$_{\pm 0.47}$ & \underline{86.56$_{\pm 1.86}$} & 80.16$_{\pm 2.40}$ & 54.35$_{\pm 7.44}$ & 59.99$_{\pm 17.31}$ & 83.72$_{\pm 1.80}$ & \textbf{87.42}$_{\bm{\pm 0.86}}$ \\
			&  & \textit{F1} $\uparrow$ & 82.07$_{\pm 0.52}$ & \underline{86.28$_{\pm 0.22}$} & 72.34$_{\pm 1.29}$ & 85.33$_{\pm 3.05}$ & 78.05$_{\pm 5.53}$ & 15.66$_{\pm 26.78}$ & 71.63$_{\pm 8.60}$ & 81.78$_{\pm 3.86}$ & \textbf{87.86}$_{\bm{\pm 0.45}}$ \\
			\cdashline{2-12}[1pt/1pt]
			& \multirow{2}{*}{\textit{256}} & \textit{Acc} $\uparrow$ & 83.53$_{\pm 0.49}$ & 86.38$_{\pm 0.12}$ & 74.71$_{\pm 0.34}$ & \underline{90.59$_{\pm 2.03}$} & 90.58$_{\pm 1.14}$ & 76.48$_{\pm 11.54}$ & 77.15$_{\pm 23.51}$ & 90.41$_{\pm 3.19}$ & \textbf{92.43}$_{\bm{\pm 1.52}}$ \\
			&  & \textit{F1} $\uparrow$ & 82.08$_{\pm 0.30}$ & 86.24$_{\pm 0.09}$ & 72.04$_{\pm 1.32}$ & 90.65$_{\pm 1.58}$ & \underline{90.79$_{\pm 0.89}$} & 70.21$_{\pm 24.17}$ & 82.60$_{\pm 13.81}$ & 90.07$_{\pm 3.66}$ & \textbf{92.39}$_{\bm{\pm 1.41}}$ \\
			\cdashline{2-12}[1pt/1pt]
			& \multirow{2}{*}{\textbf{\textit{Avg.}}} & \textit{Acc} $\uparrow$ & 83.63$_{\pm 0.73}$ & \textbf{86.41}$_{\bm{\pm 0.33}}$ & 74.54$_{\pm 0.46}$ & 79.59$_{\pm 3.52}$ & 75.59$_{\pm 2.77}$ & 57.89$_{\pm 5.01}$ & 65.99$_{\pm 17.14}$ & 76.31$_{\pm 5.79}$ & \underline{85.93$_{\pm 1.76}$} \\
			&  & \textit{F1} $\uparrow$ & 82.06$_{\pm 0.44}$ & \textbf{86.25}$_{\bm{\pm 0.22}}$ & 71.50$_{\pm 1.50}$ & 75.83$_{\pm 8.40}$ & 78.51$_{\pm 3.24}$ & 35.16$_{\pm 26.43}$ & 73.51$_{\pm 7.73}$ & 79.48$_{\pm 3.84}$ & \underline{86.02$_{\pm 1.71}$} \\
			
			\midrule
			
			\multirow{10}{*}{\textbf{SemEval}} & \multirow{2}{*}{\textit{32}} & \textit{Acc} $\uparrow$ & 84.50$_{\pm 1.06}$ & 84.02$_{\pm 1.10}$ & 79.65$_{\pm 0.23}$ & 82.76$_{\pm 1.91}$ & 61.85$_{\pm 8.74}$ & 50.63$_{\pm 1.10}$ & \underline{85.42$_{\pm 6.37}$} & 73.75$_{\pm 17.10}$ & \textbf{89.01}$_{\bm{\pm 3.53}}$ \\
			&  & \textit{F1} $\uparrow$ & 83.73$_{\pm 0.55}$ & 83.17$_{\pm 0.64}$ & 79.32$_{\pm 0.56}$ & 80.00$_{\pm 3.19}$ & 72.57$_{\pm 4.46}$ & 3.07$_{\pm 5.32}$ & \underline{87.32$_{\pm 4.71}$} & 78.99$_{\pm 10.13}$ & \textbf{89.57}$_{\bm{\pm 3.23}}$ \\
			\cdashline{2-12}[1pt/1pt]
			& \multirow{2}{*}{\textit{64}} & \textit{Acc} $\uparrow$ & 84.92$_{\pm 0.74}$ & 84.60$_{\pm 1.00}$ & 79.57$_{\pm 0.26}$ & 85.77$_{\pm 3.69}$ & 87.27$_{\pm 2.25}$ & 72.27$_{\pm 19.29}$ & \underline{89.40$_{\pm 3.37}$} & 88.82$_{\pm 2.35}$ & \textbf{90.97}$_{\bm{\pm 2.59}}$ \\
			&  & \textit{F1} $\uparrow$ & 83.93$_{\pm 0.41}$ & 83.58$_{\pm 0.65}$ & 78.95$_{\pm 1.46}$ & 84.74$_{\pm 4.80}$ & 87.06$_{\pm 1.57}$ & 78.27$_{\pm 10.22}$ & 88.80$_{\pm 4.36}$ & \underline{89.33$_{\pm 1.68}$} & \textbf{91.40}$_{\bm{\pm 2.14}}$ \\
			\cdashline{2-12}[1pt/1pt]
			& \multirow{2}{*}{\textit{128}} & \textit{Acc} $\uparrow$ & 84.85$_{\pm 0.51}$ & 84.58$_{\pm 0.82}$ & 79.70$_{\pm 0.23}$ & 90.90$_{\pm 2.79}$ & 88.90$_{\pm 3.39}$ & 86.39$_{\pm 4.03}$ & 88.09$_{\pm 6.84}$ & \textbf{92.98}$_{\bm{\pm 2.91}}$ & \underline{92.38$_{\pm 2.70}$} \\
			&  & \textit{F1} $\uparrow$ & 83.93$_{\pm 0.18}$ & 83.61$_{\pm 0.62}$ & 79.64$_{\pm 0.58}$ & 90.84$_{\pm 3.26}$ & 89.16$_{\pm 3.92}$ & 87.84$_{\pm 2.75}$ & 89.27$_{\pm 5.46}$ & \textbf{93.06}$_{\bm{\pm 2.88}}$ & \underline{92.74$_{\pm 2.29}$} \\
			\cdashline{2-12}[1pt/1pt]
			& \multirow{2}{*}{\textit{256}} & \textit{Acc} $\uparrow$ & 85.17$_{\pm 0.42}$ & 84.98$_{\pm 0.31}$ & 79.60$_{\pm 0.13}$ & 93.52$_{\pm 1.99}$ & \underline{95.28$_{\pm 0.82}$} & 91.86$_{\pm 2.26}$ & 90.90$_{\pm 2.46}$ & \textbf{95.62}$_{\bm{\pm 1.91}}$ & 95.17$_{\pm 0.68}$ \\
			&  & \textit{F1} $\uparrow$ & 84.04$_{\pm 0.09}$ & 83.87$_{\pm 0.17}$ & 79.08$_{\pm 0.24}$ & 93.76$_{\pm 1.82}$ & \underline{95.38$_{\pm 0.78}$} & 92.36$_{\pm 1.80}$ & 90.79$_{\pm 2.47}$ & \textbf{95.67}$_{\bm{\pm 1.90}}$ & 95.20$_{\pm 0.78}$ \\
			\cdashline{2-12}[1pt/1pt]
			& \multirow{2}{*}{\textbf{\textit{Avg.}}} & \textit{Acc} $\uparrow$ & 84.86$_{\pm 0.68}$ & 84.55$_{\pm 0.81}$ & 79.63$_{\pm 0.21}$ & 88.24$_{\pm 2.59}$ & 83.38$_{\pm 3.80}$ & 75.29$_{\pm 6.67}$ & \underline{88.45$_{\pm 4.76}$} & 87.79$_{\pm 6.07}$ & \textbf{91.88}$_{\bm{\pm 2.37}}$ \\
			&  & \textit{F1} $\uparrow$ & 83.91$_{\pm 0.31}$ & 83.56$_{\pm 0.52}$ & 79.25$_{\pm 0.71}$ & 87.33$_{\pm 3.27}$ & 86.04$_{\pm 2.68}$ & 65.39$_{\pm 5.02}$ & 89.04$_{\pm 4.25}$ & \underline{89.26$_{\pm 4.15}$} & \textbf{92.23}$_{\bm{\pm 2.11}}$ \\
			
			\midrule
			
			\multirow{2}{*}{\textbf{Overall}} & \multirow{2}{*}{\textbf{\textit{Avg.}}} & \textit{Acc} $\uparrow$ & 85.94$_{\pm 0.54}$ & 84.54$_{\pm 0.38}$ & 77.40$_{\pm 0.35}$ & \underline{88.41$_{\pm 3.15}$} & 84.84$_{\pm 2.35}$ & 74.46$_{\pm 5.32}$ & 85.35$_{\pm 7.88}$ & 87.23$_{\pm 4.14}$ & \textbf{93.60}$_{\bm{\pm 1.26}}$ \\
			&  & \textit{F1} $\uparrow$ & 85.02$_{\pm 0.44}$ & 83.55$_{\pm 0.31}$ & 75.51$_{\pm 1.20}$ & 87.09$_{\pm 4.57}$ & 87.06$_{\pm 2.15}$ & 61.17$_{\pm 11.75}$ & 87.76$_{\pm 5.01}$ & \underline{88.77$_{\pm 2.94}$} & \textbf{93.72}$_{\bm{\pm 1.18}}$ \\
			\bottomrule
		\end{tabular}%
	}
	\caption{\textbf{Detection performance across 4 datasets and shot sizes.} We report the mean and standard deviation of accuracy (Acc) and F1 score (F1) over 3 random seeds for each experiment (mean$_{\pm \mathrm{std}}$). $^{\dagger}$ denotes few-shot-tailored detectors, and $^{*}$ indicates adversarially trained detectors. The best result in each row is \textbf{bolded}, and the second-best is \underline{underlined}. We report TPR@FPR=1\% in Appendix~\ref{sec:tpr_fpr1_main}.}
	\label{tab:fs_mgt_results_reordered}
	\vspace{-1.5em}
\end{table*}

\subsection{Adversarial Training}
\label{AT}

We propose an adversarial training framework where the attacker and the detector are updated \emph{alternately} within each training step, enabling a co-evolution process.
Unlike prior adversarial-training approaches~\cite{hu2023radar,koike2024outfox,li2025iron}, our attacker adopts a dynamic strategy that generates adversarial examples on the fly. As training proceeds, the attacker leverages increasingly diverse texts in the retrieval pool $\mathcal{B}$ to produce more human-like adversarial rewrites, allowing the detector to learn richer humanization patterns. Meanwhile, the detector is optimized with the ACL and PBC objectives to balance accuracy and robustness in the few-shot settings.

\noindent\textbf{Training Process.}
At each training step $t$, the attacker first generates a rewrite $\tilde{X}$ of the input passage $X$ using the RAG procedure in Eq.~\eqref{eq:eq1} and Eq.~\eqref{eq:eq2}. We then update the surrogate detector $\mathcal{D}_{\mathrm{sur}}$ using the attacker loss in Eq.~\eqref{eq:att}, while keeping $\mathcal{D}_{\mathrm{tar}}$ frozen. Next, we update the target detector $\mathcal{D}_{\mathrm{tar}}$ by minimizing the detector objective in Eq.~\eqref{eq:det_total}, with $\mathcal{D}_{\mathrm{sur}}$ frozen. Finally, we add the newly generated rewrite to the retrieval pool $\mathcal{B}$, so that subsequent steps can retrieve stronger anchors and produce more challenging adversarial examples.
We detail the adversarial training process in form of pseudocode in Algorithm~\ref{alg:adv_train}, and we further analyze the training dynamics during adversarial training in Appendix~\ref{sec:training_dynamics}, demonstrating the effectiveness of our co-evolution mechanism.

\label{Adversarial Training}
\begin{figure*}[htbp]
	\centering
	\resizebox{1\textwidth}{!}{%
		\includegraphics{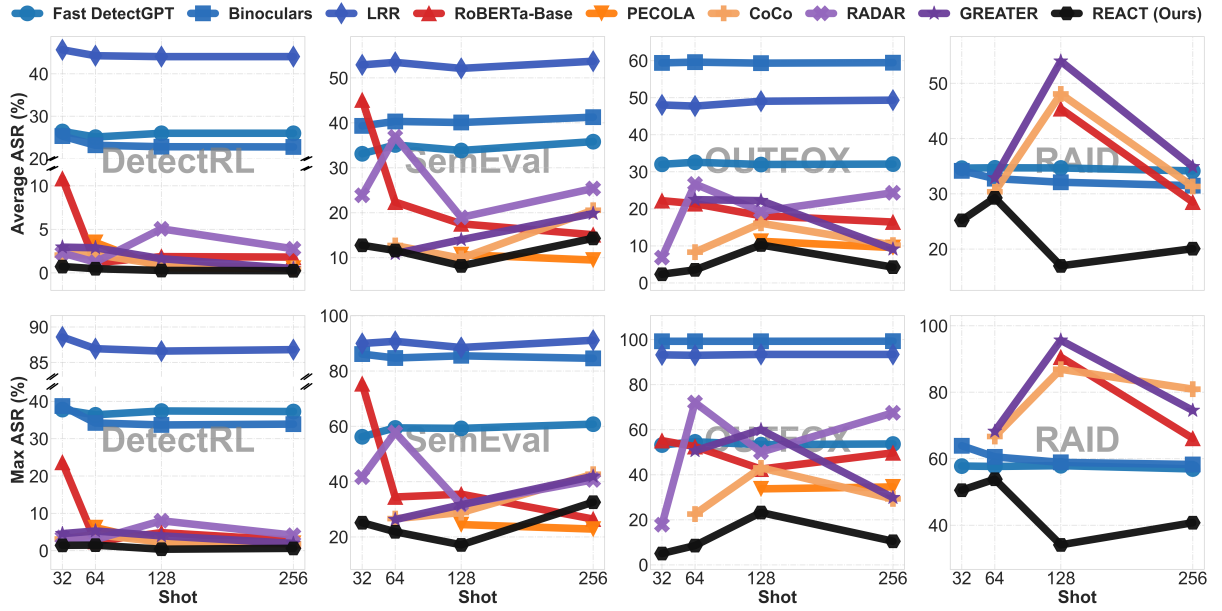} 
	}
	\caption{\textbf{Average (top) and maximum ASR (bottom) (\%) across 4 datasets and 4 shot sizes under 4 attacks.} A lower ASR (\%) indicates better model performance. We only include methods whose clean-test accuracy is above $75$ in Table~\ref{tab:fs_mgt_results_reordered}, since ASR is not informative when a detector is close to random guessing.}
	
	\label{fig:asr} 
	\vspace{-0.8em}
\end{figure*}

\section{Experiment Results}  
\label{sec:exp}

We conduct extensive experiments to comprehensively evaluate the detection accuracy of \detectorname\ in few-shot settings and its robustness under adversarial attacks. Due to space constraints, we present the ablation study results in the Appendix~\ref{sec:ablation}.



\subsection{Detection Performance in Few-shot Settings}

\fakeparagraph{Experimental setup.}
We conduct a comprehensive evaluation of \detectorname~in few-shot settings over four datasets (DetectRL~\cite{wu2024detectrl}, OUTFOX~\cite{koike2024outfox}, RAID~\cite{dugan2024raid}, and SemEval~\cite{semeval2024task8}) and four shot sizes, focusing on detection accuracy.
Our baselines include:
(i) Metric-based detectors, including Fast DetectGPT~\cite{bao2024fast}, Binoculars~\cite{hans2024spotting}, and LRR~\cite{su2023detectllm}; and
(ii) Model-based detectors, including RoBERTa-Base~\cite{liu2019roberta}, CoCo~\cite{liu2022coco}, \pecola~\cite{liu2024does}, RADAR~\cite{hu2023radar}, and \modelname~\cite{li2025iron}.
For all model-based detectors, we use the same RoBERTa-Base backbone to ensure a fair comparison.
For our method, we mainly adopt Llama-3.2-3B-Instruct~\cite{dubey2024llama} as the generator $\mathcal{G}$ and ALBERT-Base-v2~\cite{lan2019albert} as the surrogate detector $\mathcal{D}_{\mathrm{sur}}$, and the prompt template $\mathcal{Z}$ can be found in Figure~\ref{fig:prompt_template}.
To improve the robustness of our conclusions, we train each model with training sets sampled using three random seeds, and evaluate all models on the same test set.
Detailed descriptions of the baseline models and implementation details are provided in the Appendix~\ref{sec:baselines}.

\fakeparagraph{Experiment results.}
We present the few-shot detection performance of \detectorname~in Table~\ref{tab:fs_mgt_results_reordered} and unveil the following three key insights:
\textbf{1) Best overall performance.}
On the overall average, \detectorname~achieves \textbf{93.60} accuracy and \textbf{93.72} F1.
It improves over the best metric-based detector Binoculars by \textbf{+7.66} accuracy and \textbf{+8.70} F1 score, and it also surpasses the strongest model-based detector by \textbf{+5.19} accuracy and \textbf{+4.95} F1 score.
\textbf{2) Low sensitivity to random seeds.}
\detectorname~shows consistently small standard deviations under few-shot sampling.
For DetectRL with 32-shot, RADAR fluctuates substantially at $87.76_{\pm 12.49}$ accuracy, while \detectorname~remains stable at $99.46_{\pm 0.58}$ accuracy.
Similar patterns appear on OUTFOX with 32-shot, where RADAR reports $77.34_{\pm 16.01}$ accuracy but \detectorname~achieves $95.64_{\pm 1.56}$ accuracy.
\textbf{3) Stable scaling across shot sizes.}
\detectorname~is already competitive at 32-shot and scales smoothly with more supervision, especially on the challenging RAID benchmark.
Its accuracy increases from $79.30_{\pm 0.90}$ at 32-shot to $92.43_{\pm 1.52}$ at 256-shot, a gain of \textbf{+13.13} points.
At 64-shot on RAID, \detectorname~reaches $84.55_{\pm 3.78}$ accuracy, improving over the best adversarially trained detector \modelname~by \textbf{+7.34} points, and over the strongest few-shot-tailored detector CoCo by \textbf{+6.31} points.
At 128-shot, \detectorname~further surpasses \modelname~by \textbf{+3.70} points in accuracy, showing that it effectively absorbs additional data rather than being trapped in a weak few-shot solution.
Our \detectorname\ also exhibits a certain degree of generalization ability to unseen datasets, as detailed in Appendix~\ref{sec:ood}.

\subsection{Robustness against Attacks}
\label{sec:attackresult}
\fakeparagraph{Experimental setup.}
We evaluate the robustness of detectors against four strong attacks under few-shot settings on the test sets of the above four datasets. The evaluated attacks include Modify Attack~\cite{wang2024stumbling}, Paraphrasing~\cite{krishna2023paraphrasing}, Back-translation~\cite{sennrich2015improving}, and Humanize Attack~\cite{wang2024stumbling}.
Detailed descriptions of these attack methods are provided in the Appendix~\ref{sec:attack_methods}.

\fakeparagraph{Experiment results.}
We report robustness against four strong attacks in Figure~\ref{fig:asr} and summarize three findings.
\textbf{1) Lowest average ASR in most settings.}
\detectorname~achieves the lowest average ASR in $14$ out of $16$ dataset--shot settings.
Averaged over all datasets and shot sizes, it achieves \textbf{10.04\%} average ASR and reduces average ASR by \textbf{3.66} percentage points compared to the best-performing baseline in each dataset--shot setting, with larger gains at low shots (e.g., \textbf{6.47} percentage points on average at $32$-shot).
\textbf{2) Strongest worst-case robustness under max ASR.}
\detectorname~also attains the lowest max ASR in $15$ out of $16$ settings.
Across all datasets and shot sizes, it lowers max ASR by \textbf{8.20} percentage points on average, with especially large margins on challenging cases such as OUTFOX at $256$-shot (\textbf{18.77} percentage points) and RAID at $128$-shot (\textbf{23.81} percentage points).
\textbf{3) Consistent gains across shot sizes.}
The reductions remain positive at every shot size.
For average ASR, the mean gap is \textbf{6.47}, \textbf{1.46}, \textbf{4.52}, and \textbf{2.18} percentage points at $32$, $64$, $128$, and $256$ shots, respectively; the same pattern holds for max ASR with \textbf{9.56}, \textbf{5.80}, \textbf{10.84}, and \textbf{6.60} percentage points, indicating that \detectorname~provides stable robustness improvements from scarce supervision to higher-shot settings.

\vspace{-0.2cm}

\section{Discussion}
\label{sec:discussion}


\subsection{Impact of Adversarial Example Generation Strategy}
\label{sec:generation}
\begin{figure}[t]
    \centering
    \resizebox{0.48\textwidth}{!}{    \includegraphics{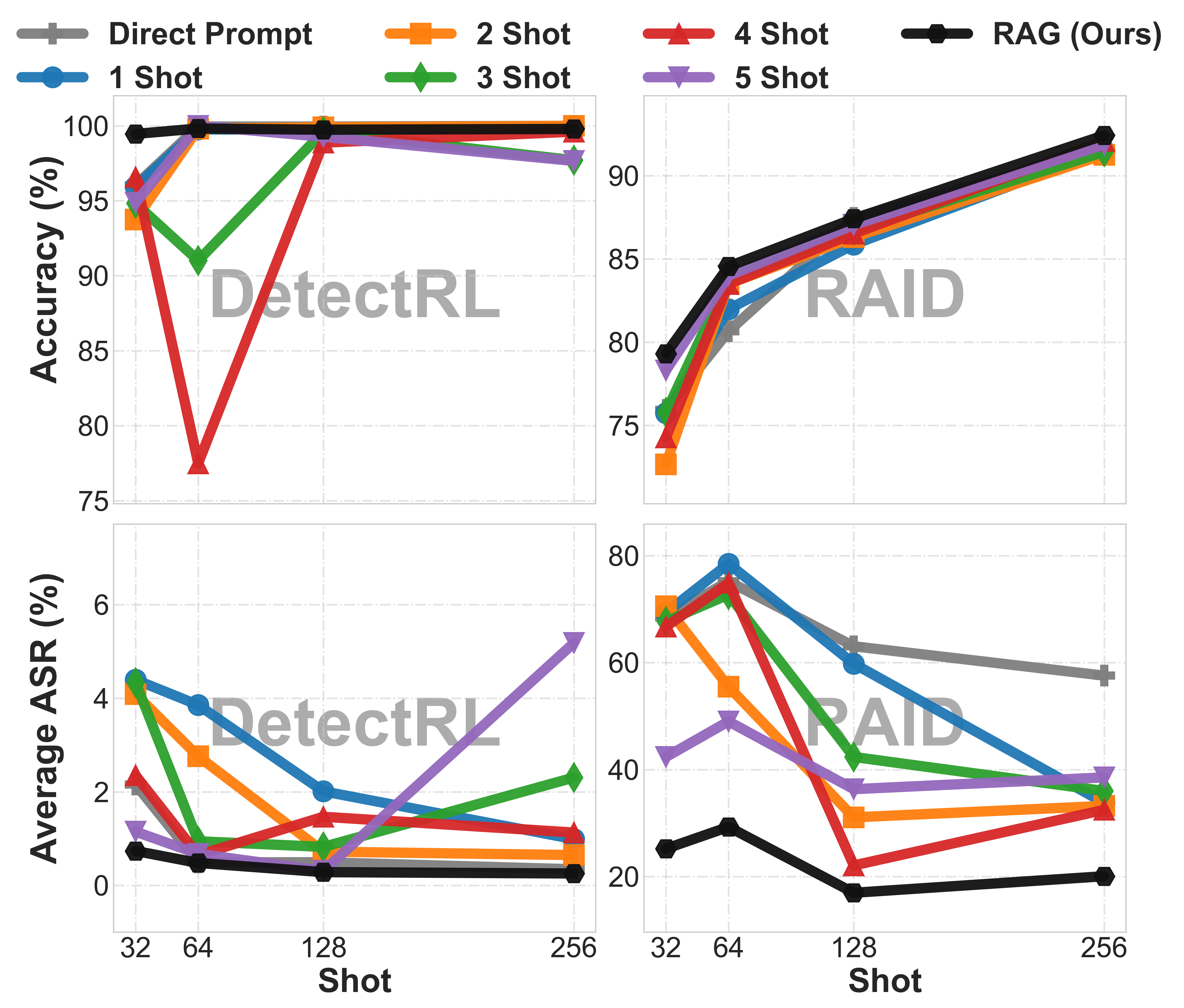} 
    }
	\caption{
    \textbf{Impact of adversarial example generation strategy in \detectorname.}
    A higher accuracy (top) and lower ASR (bottom) indicates higher model performance.
	}
    \label{fig:compare} 
\end{figure}

In this section, we study how the adversarial example generation strategy affects \detectorname's accuracy and robustness.
As described in \secref{sec:attacker}, our attacker adopts RAG, while other schemes are alternative prompting baselines.
We compare RAG with two prompt-based baselines: \textit{Direct Prompt}, which uses a fixed template to instruct the generator to produce human-like rewrites, and \textit{Few-shot Prompting}, which samples $k\!\in\!\{1,2,3,4,5\}$ $(\mathrm{MGT},\mathrm{HWT})$ pairs as in-context exemplars.
We evaluate these strategies on DetectRL and RAID, reporting clean accuracy and ASR under four attacks across shot sizes.
Results shown in Figure~\ref{fig:compare} yield two findings.
\textbf{First, RAG offers the best accuracy--robustness trade-off with minimal context.}
With only one retrieved pair, RAG achieves higher accuracy and lower ASR than all prompt-based baselines, while reducing token budget and training cost.
\textbf{Second, more in-context shots can help but reduce stability.}
Although larger $k$ sometimes lowers ASR, prompt-based baselines become less consistent; for example, 4-shot prompting reaches $\sim$20\% ASR on RAID at 128-shot but rises to nearly 70\% at 64-shot.
In contrast, RAG remains consistently strong across settings.

\vspace{-0.1cm}
\section{Conclusion}
In this paper, we propose an adversarial training framework named
\textbf{R}AG-Guid\textbf{E}d
\textbf{A}ttacker Strengthens \textbf{C}on\textbf{T}rastive
Few-shot Detector (\detectorname) to improve the detection accuracy of MGT detectors in few-shot settings as well as their robustness against attacks.
We design a novel RAG-based adversarial example generation strategy for the attacker and introduce a pairwise boundary contrastive loss for the detector to enhance few-shot detection performance.
Extensive experiments on 4 datasets with 4 shot sizes and 3 random seeds demonstrate that \detectorname~consistently outperforms 8 SOTA detectors in both detection accuracy and robustness.
Further discussion of adversarial example generation strategies show that RAG achieves better training effectiveness while reducing training time compared with alternative few-shot prompting methods.



\section*{Limitations}

Despite the strong performance of \detectorname, it still has several limitations.
\textbf{First}, while \detectorname~performs well on English benchmarks, extending it to multilingual settings remains challenging.
\textbf{Second}, although adversarial training yields clear improvements, it also increases training time and incurs additional computational and resource overhead during training.
\textbf{Third}, the detector in our work operates at the document level and therefore cannot localize machine-generated content within a human--AI co-authored text. Developing fine-grained attribution methods, such as sentence-level or token-level detection, remains an important direction for future work.

\section*{Ethics Statement}

This work studies and improves the robustness of machine-generated text (MGT) detection under strong humanization-style attacks, with a particular focus on few-shot settings. Our goal is defensive: we design an adversarial training framework that helps detectors learn more reliable, attack-resilient decision boundaries, rather than enabling content deception in real-world systems.

\paragraph{Dual-use considerations.}
Techniques that generate more human-like rewrites are inherently dual-use: they could be misused to evade detectors, facilitate plagiarism, or amplify misinformation. We therefore frame the attacker strictly as a \emph{training-time} component to harden detectors. Our experiments and analysis are intended to expose vulnerabilities and improve defensive robustness, not to provide practical evasion guidance.

\paragraph{Prompt disclosure and risk mitigation.}
For transparency and reproducibility, we include a prompt \emph{template} in the appendix. However, we take care to avoid releasing a turnkey evasion recipe. In particular, (i) we do not provide an automated end-to-end evasion pipeline (e.g., query-efficient search loops, attack orchestration code, or deployment-ready scripts), (ii) we do not include systematically optimized prompts, prompt ensembles, or model-specific prompt tuning details that would materially lower the barrier to misuse, and (iii) we emphasize that effectiveness in our framework relies on the \emph{full} adversarial-training setup (e.g., retrieval pool maintenance and surrogate-aligned scoring under a constrained black-box feedback signal), rather than the prompt alone. We strongly discourage using any components of this work to bypass safety or provenance mechanisms.

\paragraph{Responsible release and intended use.}
Any artifacts we release are intended to support \emph{defensive} research and reproducibility. We encourage practitioners to combine detection with appropriate safeguards (e.g., human-in-the-loop review, transparency mechanisms, and due-process policies) and to consider rate limiting and abuse monitoring when detectors are deployed as public APIs.

\paragraph{Data, privacy, and content safety.}
All experiments are conducted on publicly available benchmark datasets. We do not collect new user data and do not target private, personally identifiable, or sensitive information. Generated rewrites are used solely for evaluating robustness and training defenses, and are not used for surveillance or content censorship.

\paragraph{Human evaluation.}
When human evaluation is performed (Appendix~\ref{sec:human}), we follow standard annotation protocols to minimize risk to annotators, avoid exposing private information, and restrict the task to assessing writing quality and human-likeness. Annotators are informed of the task purpose and the data sources, and the study is conducted in accordance with applicable institutional and legal requirements.

\paragraph{Environmental and resource impact.}
Adversarial training increases compute cost due to iterative generation and alternating updates. We mitigate this by using a controlled experimental setup and reporting results under fixed budgets. Future work should further reduce training overhead while preserving robustness gains.

Overall, we present \detectorname\ as a defensive contribution to strengthening MGT detection under realistic attack pressures, and we discourage any misuse of the methodology or disclosed materials for deceptive purposes.


\bibliography{GENSHIN}     

\clearpage
\newpage
\appendix
\section{Experimental Setting}
\label{apdx:expset}


\subsection{Dataset}
\label{sec:dataset}

We conduct experiments on four benchmark datasets: DetectRL~\cite{wu2024detectrl}, OUTFOX~\cite{koike2024outfox}, RAID~\cite{dugan2024raid}, and SemEval~\cite{semeval2024task8}. 
For each dataset, we construct few-shot training sets by sampling $\{32,64,128,256\}$ labeled instances using three random seeds $\{66,2025,9999\}$. 
For evaluation, we sample 2{,}000 \emph{disjoint} instances (1{,}000 HWT + 1{,}000 MGT) as a balanced test set from the corresponding dataset split, ensuring no overlap with the few-shot training set.
We additionally use HC3~\cite{guo2023close} for OOD evaluation.

\subsection{Implementation}
\label{sec:implementation}

\detectorname\ is deployed on a server equipped with two RTX 4090 GPUs running Ubuntu 22.04.
Our adversarial training framework mainly adopts Llama-3.2-3B-Instruct~\cite{dubey2024llama} as the generator $\mathcal{G}$ and ALBERT-Base-v2~\cite{lan2019albert} as the surrogate detector $\mathcal{D}_{\mathrm{sur}}$.
We use RoBERTa-Base~\cite{liu2019roberta} as the backbone of the target detector $\mathcal{D}_{\mathrm{tar}}$.
For metric-based detectors, we train a logistic-regression calibrator on the few-shot training set to map raw scores to calibrated probabilities (of the MGT class), and report Acc/F1 using a fixed decision threshold of 0.5.
For model-based detectors, we employ the same RoBERTa-Base backbone to ensure a fair comparison, train each method for 6 epochs under an identical protocol, and report results using the final checkpoint with a fixed decision threshold of 0.5.
Under this 6-epoch protocol, \detectorname\ uses a two-stage schedule: we first pre-train on clean samples using only the cross-entropy loss for 3 epochs, and then perform adversarial training for another 3 epochs.
The selected hyperparameters of \detectorname\ are summarized in Table~\ref{tab:hyperparameters}.

\begin{table}[h]
    \centering
    \resizebox{0.45\textwidth}{!}{ 
        \begin{tabular}{cc}
            \toprule
            Hyperparameter & Value \\
            \midrule
            Pre-training epochs $epoch_{\mathrm{pre}}$ & 3 \\
            Maximum adversarial epochs $epoch_{\max}$ & 3 \\
            Total epochs & 6 \\
            
            Target detector learning rate $\eta_{\mathrm{tar}}$ & $2\times 10^{-5}$ \\
            Surrogate detector learning rate $\eta_{\mathrm{sur}}$ & $2\times 10^{-5}$ \\
            AdamW weight decay $\beta$ & 0.03 \\
            
            Batch size $M$ & 2 \\
            
            Max sequence length $L$ (for $\mathcal{D}_{\mathrm{tar}},\mathcal{D}_{\mathrm{sur}}$) & 512 \\
            
            Adversarial CE scaling coefficient $\alpha$ & 0.5 \\
            PBC loss weight $\lambda_{\mathrm{pbc}}$ & 1.2 \\
            PBC margin for same-class $\delta_{\mathrm{same}}$ & 0.1 \\
            PBC margin for diff-class $\delta_{\mathrm{diff}}$ & 0.3 \\
            
            Generator max new tokens $T_{\mathrm{new}}$ & 512 \\
            Generator input max length $L_{\mathrm{gen}}$ & 4096 \\
            Sampling temperature $\tau$ & 0.7 \\
            Nucleus sampling $p$ & 0.9 \\          
            \bottomrule
        \end{tabular}
    }
    \caption{Hyperparameters for our \detectorname.}
    \label{tab:hyperparameters}
\end{table}

We use the following prompt template $\mathcal{Z}(\cdot)$ in Figure~\ref{fig:prompt_template}:

\begin{figure*}[t]
	\centering
	\footnotesize
	\fbox{%
		\begin{minipage}{1\linewidth}
			\textbf{Prompt template $\mathcal{Z}(X;X^{H},X^{M})$}\\
			\vspace{2pt}
			\textbf{Human-style reference:}
			\texttt{$X^{H}$}\\
			\vspace{2pt}
			\textbf{Machine-style reference:}
			\texttt{$X^{M}$}\\
			\vspace{2pt}
			\textbf{Target passage:}
			\texttt{$X$}\\
			\vspace{4pt}
			\textbf{Instruction:}\\
			You are a de-AI-style rewriter. Your task is to rewrite a given text so that it avoids being detected as AI-generated. \\
			Your objectives: \\
			- Preserve the original meaning, facts, and intent of the text. \\
			- Make the rewritten text look as if it were written by a human. \\
			- Strictly write in the SAME LANGUAGE as the TARGET\_TEXT. \\
			- Your maximum generation length is \{max\_tokens\} tokens. \\
			You will be given two reference texts: \\
			1. HUMAN\_LIKE\_REFERENCE: a text (or several examples) that the detector considers the most human-like. \\
			2. AI\_LIKE\_REFERENCE: a text (or several examples) that the detector considers the most AI-like. \\
			Your style constraints: \\
			- Imitate the style, tone, and rhythm of HUMAN\_LIKE\_REFERENCE. \\
			- Explicitly avoid any stylistic patterns that resemble AI\_LIKE\_REFERENCE. \\
			- Do NOT mention detectors, AI, models, prompts, or the rewriting process. \\
			- Do NOT add explanations, comments, or meta-text. Output ONLY the rewritten text. \\
			Steps you MUST follow internally (do NOT output the steps): \\
			1. Carefully read HUMAN\_LIKE\_REFERENCE and understand its style and tone. \\
			2. Briefly compare it in your mind with AI\_LIKE\_REFERENCE and identify stylistic differences. \\
			3. Rewrite TARGET\_TEXT entirely in the style of HUMAN\_LIKE\_REFERENCE, while avoiding the style of AI-LIKE\_REFERENCE. \\
			Now here are the references and the target: \\
			HUMAN\_LIKE\_REFERENCE: \\
			$X^{H}$ \\
			AI\_LIKE\_REFERENCE: \\
			$X^{M}$ \\
			TARGET\_TEXT: \\
			$X$ \\
			Rewritten TARGET\_TEXT (remember: output ONLY the rewritten text): 
		\end{minipage}%
	}
	\caption{Prompt template used to construct $\mathcal{Z}(X;X^{H},X^{M})$ for RAG-based rewriting.}
	\label{fig:prompt_template}
\end{figure*}

\subsection{Evaluation Metrics}
\label{apdx:metric}
To evaluate detection performance, we report \emph{accuracy} (Acc) and \emph{F1-score} (F1).
To quantify robustness against adversarial attacks, we use the \emph{attack success rate} (ASR), where a lower ASR indicates stronger robustness.
Under our threat model, the attacker perturbs only machine-generated texts (MGTs).
Accordingly, ASR is defined as the fraction of originally correctly classified MGTs that become misclassified after the attack:
\begin{equation}
	\label{eq:asr}
	\small
	\mathrm{ASR}
	=
	\frac{
		\#\{\text{MGTs misclassified after attack}\}
	}{
		\#\{\text{MGTs correctly classified before attack}\}
	}.
\end{equation}

\subsection{Experimental Comparison Methods}
\label{sec:baselines}

To better assess the effectiveness of \detectorname, we compare it with representative state-of-the-art (SOTA) MGT detectors, including three widely used \emph{metric-based} detectors and five \emph{model-based} detectors.
Metric-based detectors typically output \emph{raw scores} rather than calibrated probabilities.
Following the standard practice in their original papers, we perform score calibration on the corresponding training set by fitting a logistic regression model that maps each detector's raw score to an output probability for binary classification.
For each metric-based detector, we follow the backbone choices recommended by its original paper (and official implementation) for the underlying language model(s).
In particular, for methods that require an LM backbone, we use the paper-recommended scoring/sampling model for Fast DetectGPT, the recommended performer/observer pair for Binoculars, and the recommended scoring model for LRR.

\begin{itemize}
	\item \textbf{Fast DetectGPT}~\cite{bao2024fast}:
	A metric-based zero-shot detector derived from DetectGPT, designed to reduce the computational overhead of perturbation-based scoring.
	Instead of repeatedly perturbing the input, it measures \emph{conditional probability curvature}, which captures discrepancies in token-level likelihood geometry between human-written and machine-generated text.
	This single-pass scoring strategy substantially accelerates inference while maintaining competitive detection quality.
	
	\item \textbf{Binoculars}~\cite{hans2024spotting}:
	A metric-based zero-shot detector that detects machine-generated text by contrasting the output distributions of two closely related pretrained language models.
	It computes a lightweight divergence-style score that reflects subtle inconsistencies between human and machine writing, requiring neither fine-tuning nor labeled data.
	
	\item \textbf{LRR}~\cite{su2023detectllm}:
	A metric-based zero-shot detector that uses log-rank statistics under a reference language model.
	It identifies atypical token-rank patterns that commonly arise in LLM generations, offering a fast alternative to perturbation-heavy methods while retaining strong empirical performance.
	
	\item \textbf{RoBERTa-Base}~\cite{liu2019roberta}:
	A standard supervised detector that fine-tunes an RoBERTa-Base classifier on the labeled training set and directly outputs class probabilities for binary MGT detection.
	
	\item \textbf{CoCo}~\cite{liu2022coco}:
	A few-shot-tailored detector that introduces contrastive training driven by \emph{local coherence} signals.
	It first scores each input using local coherence computed by a set of in-domain language models, and then trains the detector with a contrastive objective to better separate human-written texts and machine-generated texts in representation space.
	Since CoCo is primarily designed for sample-efficient discrimination rather than adversarial robustness, it can be less robust under strong attacks in our robustness evaluation.
	
	\item \textbf{\pecola}~\cite{liu2024does}:
	A contrastive-learning-based detector that improves fine-tuned MGT detectors via \emph{selective perturbation} and a contrastive objective to learn more discriminative representations.
	We treat \pecola\ as a few-shot-tailored baseline due to its sample-efficient training design, but it is not explicitly adversarially trained, which can lead to weaker robustness under our attack suite.
	
	\item \textbf{RADAR}~\cite{hu2023radar}:
	An adversarially trained detector that improves robustness against paraphrasing-style attacks by training the detector jointly with a paraphraser.
	The paraphraser generates low-perplexity paraphrases to evade detection, while the detector learns to correctly classify both original texts and adversarial paraphrases.
	RADAR focuses on monolingual paraphrase-based adversaries and is mainly proposed as a robustness-oriented training recipe; thus, it is not specifically tailored to few-shot training settings.
	
	\item \textbf{\modelname}~\cite{li2025iron}:
	An adversarially trained detector within the GREATER framework, where an adversary identifies and perturbs critical tokens and searches for effective adversarial examples.
	\modelname\ updates the adversary and the detector synchronously so that the detector learns to defend against evolving attacks and generalize to other perturbations.
	As a robustness-first adversarial training framework, \modelname\ is not originally designed for few-shot settings and typically incurs higher training overhead due to the coupled attacker--defender optimization.
\end{itemize}

\subsection{Detailed Information of Attack Method}
\label{sec:attack_methods}

Here we provide detailed descriptions of the 4 strong attack methods used in the main paper.

\begin{itemize}
	\item \textbf{Modify Attack}~\cite{wang2024stumbling}:
	A mask-filling style humanization attack implemented with \textbf{T5-large}~\cite{2020t5}.
	Given an input MGT, we mask a subset of tokens and let T5-large fill in the blanks, producing a rewritten text that largely preserves the original semantics while injecting more human-like lexical and syntactic variations.
	This attack targets detectors that rely on surface cues or token-level likelihood patterns by introducing semantics-preserving perturbations.
	
	\item \textbf{Paraphrasing}~\cite{krishna2023paraphrasing}:
	A sentence-level paraphrase attack implemented with \textbf{DIPPER}~\cite{krishna2023paraphrasing}.
	For each MGT input, DIPPER rewrites the passage to preserve meaning while substantially altering phrasing and local syntax, yielding human-like paraphrases.
	This attack aims to evade detection via semantic camouflage by replacing characteristic generation artifacts with more natural expressions.
	
	\item \textbf{Back-translation}~\cite{sennrich2015improving}:
	A sentence-level paraphrase attack that round-trip translates each MGT through a pivot language to induce lexical and syntactic diversity while preserving meaning.
	We use public \textbf{Helsinki-NLP OPUS-MT} models~\cite{TiedemannThottingal:EAMT2020} and choose \textbf{German} as the pivot language.
	Concretely, we translate the input text into German and then back to the original language, and replace the original text with the back-translated output.
	
	\item \textbf{Humanize Attack}~\cite{wang2024stumbling}:
	An in-context learning based humanization attack that uses \textbf{three} $(\mathrm{MGT},\mathrm{HWT})$ pairs as in-context exemplars to guide rewriting.
	Given an MGT input, the attacker prompts the generator with these paired exemplars to synthesize a highly human-like rewrite while keeping the underlying meaning.
	We use \textbf{DeepSeek V3.2}~\cite{liu2025deepseek} as the generator for this attack.
\end{itemize}

\section{Additional Result}

\subsection{Ablation Study}
\label{sec:ablation}
\fakeparagraph{Experiment Setup.}To verify the role of each component in \detectorname, we conduct an ablation study that strictly follows the same experimental setting as Table~\ref{tab:fs_mgt_results_reordered} so that the conclusions are directly comparable. 
Under this controlled setup, we compare four variants in Table~\ref{tab:fs_acc_avg_asr_partial_with_worag_recalc}: a standard supervised \textbf{RoBERTa-Base} detector, a variant that removes the pairwise boundary-constraint loss $\mathcal{L}_{\mathrm{PBC}}$ from Eq.~\eqref{eq:det_total} (\textbf{w/o $\mathcal{L}_{\mathrm{PBC}}$}), a variant that disables RAG and instead uses direct prompting to generate adversarial rewrites (\textbf{w/o RAG}), and the full \textbf{\detectorname}. 
\textbf{We do not include the ablation that removes $\mathcal{L}_{\mathrm{ACL}}$, because $\mathcal{L}_{\mathrm{ACL}}$ is the classification loss; without it, the detector cannot be trained to perform the classification task.}

\begin{table}[t]
	\centering
	\renewcommand\arraystretch{1.2}
	\footnotesize
	\resizebox{0.45\textwidth}{!}{%
		\begin{tabular}{c c c c c c c}
			\toprule
			\textbf{Dataset} & \textbf{Shot} & \textbf{Metric} & \textbf{RoBERTa-Base} & \textbf{w/o $\mathcal{L}_{\mathrm{PBC}}$} & \textbf{w/o RAG} & \textbf{\detectorname~(Ours)} \\
			\midrule
			
			\multirow{10}{*}{\textbf{DetectRL}} & \multirow{2}{*}{\textit{32}} & \textit{Acc} $\uparrow$
			& 86.08$_{\pm 0.96}$ & \underline{96.89$_{\pm 4.88}$} & 96.29$_{\pm 2.09}$ & \textbf{99.46$_{\pm 0.58}$} \\
			&  & \textit{Avg ASR} $\downarrow$
			& 10.84 & \underline{1.36} & 2.04 & \textbf{0.74} \\
			\cdashline{2-7}[1pt/1pt]
			
			& \multirow{2}{*}{\textit{64}} & \textit{Acc} $\uparrow$
			& 99.43$_{\pm 0.61}$ & \underline{99.77$_{\pm 0.06}$} & 99.72$_{\pm 0.10}$ & \textbf{99.82$_{\pm 0.27}$} \\
			&  & \textit{Avg ASR} $\downarrow$
			& 1.16 & \underline{0.60} & 0.81 & \textbf{0.48} \\
			\cdashline{2-7}[1pt/1pt]
			
			& \multirow{2}{*}{\textit{128}} & \textit{Acc} $\uparrow$
			& 95.54$_{\pm 6.69}$ & \underline{99.72$_{\pm 0.07}$} & 99.71$_{\pm 0.05}$ & \textbf{99.74$_{\pm 0.07}$} \\
			&  & \textit{Avg ASR} $\downarrow$
			& 1.87 & 0.46 & \underline{0.44} & \textbf{0.28} \\
			\cdashline{2-7}[1pt/1pt]
			
			& \multirow{2}{*}{\textit{256}} & \textit{Acc} $\uparrow$
			& 98.08$_{\pm 2.78}$ & \textbf{99.79$_{\pm 0.16}$} & \underline{99.77$_{\pm 0.06}$} & \textbf{99.79$_{\pm 0.16}$} \\
			&  & \textit{Avg ASR} $\downarrow$
			& 1.82 & 0.43 & \underline{0.38} & \textbf{0.26} \\
			\cdashline{2-7}[1pt/1pt]
			
			& \multirow{2}{*}{\textbf{\textit{Avg.}}} & \textit{Acc} $\uparrow$
			& 94.78$_{\pm 2.76}$ & \underline{99.04$_{\pm 1.29}$} & 98.87$_{\pm 0.58}$ & \textbf{99.70$_{\pm 0.27}$} \\
			&  & \textit{Avg ASR} $\downarrow$
			& 3.92 & \underline{0.71} & 0.92 & \textbf{0.44} \\
			
			\midrule
			
			\multirow{10}{*}{\textbf{OUTFOX}} & \multirow{2}{*}{\textit{32}} & \textit{Acc} $\uparrow$
			& 88.15$_{\pm 2.78}$ & 91.72$_{\pm 1.76}$ & \underline{92.20$_{\pm 3.17}$} & \textbf{95.64$_{\pm 1.56}$} \\
			&  & \textit{Avg ASR} $\downarrow$
			& 22.20 & \underline{13.79} & 15.66 & \textbf{2.36} \\
			\cdashline{2-7}[1pt/1pt]
			
			& \multirow{2}{*}{\textit{64}} & \textit{Acc} $\uparrow$
			& 90.84$_{\pm 1.74}$ & 93.12$_{\pm 6.09}$ & \underline{94.19$_{\pm 3.64}$} & \textbf{97.12$_{\pm 0.05}$} \\
			&  & \textit{Avg ASR} $\downarrow$
			& 21.39 & \underline{13.09} & 14.59 & \textbf{3.53} \\
			\cdashline{2-7}[1pt/1pt]
			
			& \multirow{2}{*}{\textit{128}} & \textit{Acc} $\uparrow$
			& 87.65$_{\pm 9.64}$ & 95.41$_{\pm 0.35}$ & \underline{96.18$_{\pm 2.65}$} & \textbf{96.73$_{\pm 0.48}$} \\
			&  & \textit{Avg ASR} $\downarrow$
			& 18.14 & \underline{14.82} & 15.52 & \textbf{10.26} \\
			\cdashline{2-7}[1pt/1pt]
			
			& \multirow{2}{*}{\textit{256}} & \textit{Acc} $\uparrow$
			& \underline{97.56$_{\pm 0.82}$} & 97.03$_{\pm 0.44}$ & 97.40$_{\pm 0.34}$ & \textbf{98.03$_{\pm 0.52}$} \\
			&  & \textit{Avg ASR} $\downarrow$
			& 16.43 & 4.92 & \underline{4.87} & \textbf{4.27} \\
			\cdashline{2-7}[1pt/1pt]
			
			& \multirow{2}{*}{\textbf{\textit{Avg.}}} & \textit{Acc} $\uparrow$
			& 91.05$_{\pm 3.74}$ & 94.32$_{\pm 2.16}$ & \underline{94.99$_{\pm 2.45}$} & \textbf{96.88$_{\pm 0.65}$} \\
			&  & \textit{Avg ASR} $\downarrow$
			& 19.54 & \underline{11.66} & 12.66 & \textbf{5.11} \\
			
			\midrule
			
			\multirow{10}{*}{\textbf{RAID}} & \multirow{2}{*}{\textit{32}} & \textit{Acc} $\uparrow$
			& 72.09$_{\pm 4.59}$ & \underline{76.35$_{\pm 7.45}$} & 75.08$_{\pm 0.96}$ & \textbf{79.30$_{\pm 0.90}$} \\
			&  & \textit{Avg ASR} $\downarrow$
			& -- & 72.72 & \underline{65.14} & \textbf{25.21} \\
			\cdashline{2-7}[1pt/1pt]
			
			& \multirow{2}{*}{\textit{64}} & \textit{Acc} $\uparrow$
			& 69.12$_{\pm 6.59}$ & 77.43$_{\pm 1.95}$ & \underline{80.78$_{\pm 1.27}$} & \textbf{84.55$_{\pm 3.78}$} \\
			&  & \textit{Avg ASR} $\downarrow$
			& -- & 83.90 & \underline{74.26} & \textbf{29.28} \\
			\cdashline{2-7}[1pt/1pt]
			
			& \multirow{2}{*}{\textit{128}} & \textit{Acc} $\uparrow$
			& 86.56$_{\pm 1.86}$ & 86.87$_{\pm 3.30}$ & \underline{86.91$_{\pm 3.19}$} & \textbf{87.42$_{\pm 0.86}$} \\
			&  & \textit{Avg ASR} $\downarrow$
			& 45.41 & \underline{42.45} & 61.78 & \textbf{16.97} \\
			\cdashline{2-7}[1pt/1pt]
			
			& \multirow{2}{*}{\textit{256}} & \textit{Acc} $\uparrow$
			& 90.59$_{\pm 2.03}$ & 91.01$_{\pm 3.83}$ & \underline{92.29$_{\pm 0.98}$} & \textbf{92.43$_{\pm 1.52}$} \\
			&  & \textit{Avg ASR} $\downarrow$
			& 28.51 & \underline{26.49} & 61.65 & \textbf{20.07} \\
			\cdashline{2-7}[1pt/1pt]
			
			& \multirow{2}{*}{\textbf{\textit{Avg.}}} & \textit{Acc} $\uparrow$
			& 79.59$_{\pm 3.77}$ & 82.92$_{\pm 4.13}$ & \underline{83.77$_{\pm 1.60}$} & \textbf{85.93$_{\pm 1.77}$} \\
			&  & \textit{Avg ASR} $\downarrow$
			& \underline{36.96} & 56.39 & 65.71 & \textbf{22.89} \\
			
			\midrule
			
			\multirow{10}{*}{\textbf{SemEval}} & \multirow{2}{*}{\textit{32}} & \textit{Acc} $\uparrow$
			& 82.76$_{\pm 1.91}$ & \underline{85.60$_{\pm 5.29}$} & 79.13$_{\pm 7.85}$ & \textbf{89.01$_{\pm 3.53}$} \\
			&  & \textit{Avg ASR} $\downarrow$
			& 45.09 & 17.08 & \underline{16.65} & \textbf{12.77} \\
			\cdashline{2-7}[1pt/1pt]
			
			& \multirow{2}{*}{\textit{64}} & \textit{Acc} $\uparrow$
			& 85.77$_{\pm 3.69}$ & 87.65$_{\pm 2.28}$ & \underline{89.37$_{\pm 6.35}$} & \textbf{90.97$_{\pm 2.59}$} \\
			&  & \textit{Avg ASR} $\downarrow$
			& 22.38 & \underline{13.98} & 14.98 & \textbf{11.60} \\
			\cdashline{2-7}[1pt/1pt]
			
			& \multirow{2}{*}{\textit{128}} & \textit{Acc} $\uparrow$
			& 90.90$_{\pm 2.79}$ & \underline{91.62$_{\pm 4.56}$} & 89.29$_{\pm 4.62}$ & \textbf{92.38$_{\pm 2.70}$} \\
			&  & \textit{Avg ASR} $\downarrow$
			& 17.53 & \underline{12.43} & 13.21 & \textbf{8.19} \\
			\cdashline{2-7}[1pt/1pt]
			
			& \multirow{2}{*}{\textit{256}} & \textit{Acc} $\uparrow$
			& 93.52$_{\pm 1.99}$ & 93.95$_{\pm 1.03}$ & \underline{94.43$_{\pm 3.66}$} & \textbf{95.17$_{\pm 0.68}$} \\
			&  & \textit{Avg ASR} $\downarrow$
			& 15.10 & 14.86 & \underline{14.49} & \textbf{14.30} \\
			\cdashline{2-7}[1pt/1pt]
			
			& \multirow{2}{*}{\textbf{\textit{Avg.}}} & \textit{Acc} $\uparrow$
			& 88.24$_{\pm 2.59}$ & \underline{89.70$_{\pm 3.29}$} & 88.06$_{\pm 5.62}$ & \textbf{91.88$_{\pm 2.37}$} \\
			&  & \textit{Avg ASR} $\downarrow$
			& 25.03 & \underline{14.59} & 14.83 & \textbf{11.72} \\
			
			\midrule
			
			\multirow{2}{*}{\textbf{Overall}} & \multirow{2}{*}{\textbf{\textit{Avg.}}} & \textit{Acc} $\uparrow$
			& 88.41$_{\pm 3.22}$ & \underline{91.50$_{\pm 2.72}$} & 91.42$_{\pm 2.56}$ & \textbf{93.60$_{\pm 1.26}$} \\
			&  & \textit{Avg ASR} $\downarrow$
			& 21.36 & \underline{20.84} & 23.53 & \textbf{10.04} \\
			
			\bottomrule
		\end{tabular}%
	}
	\caption{\textbf{Result of Ablation Study.} Accuracy (Acc, \%) is reported as mean$_{\pm \mathrm{std}}$ over three random seeds. Avg ASR (\%) denotes the average attack success rate over four attacks. The best result in each row is \textbf{bolded}, and the second-best is \underline{underlined}.}
	\label{tab:fs_acc_avg_asr_partial_with_worag_recalc}
\end{table}

\fakeparagraph{Experiment results.}
We report the ablation results in Table~\ref{tab:fs_acc_avg_asr_partial_with_worag_recalc} and summarize three findings.
Following Table~\ref{tab:fs_mgt_results_reordered}, we report ASR only for detectors whose clean-test accuracy exceeds $75\%$, since ASR is not informative when a detector is close to random guessing (hence ``--'' for RoBERTa-Base on RAID at 32/64 shots).

\textbf{1) RAG is a major driver of robustness by strengthening humanization-oriented adversarial rewrites.}
Disabling retrieval guidance (\textbf{w/o RAG}) substantially weakens adversarial training, increasing the overall Avg ASR from \textbf{10.04} to \textbf{23.53} and reducing accuracy from \textbf{93.60} to \textbf{91.42}.
The effect is most pronounced on RAID, where Avg ASR rises from \textbf{22.89} to \textbf{65.71}; at 128-shot it jumps from \textbf{16.97} to \textbf{61.78}, highlighting the importance of strong adversarial rewrites for robustness.

\textbf{2) $\mathcal{L}_{\mathrm{PBC}}$ improves the accuracy--robustness trade-off and stabilizes few-shot training.}
Removing the pairwise boundary constraint (\textbf{w/o $\mathcal{L}_{\mathrm{PBC}}$}) degrades performance on average, reducing the overall accuracy from \textbf{93.60} to \textbf{91.50} and increasing the overall Avg ASR from \textbf{10.04} to \textbf{20.84}.
On RAID, Avg ASR further deteriorates to \textbf{56.39}, suggesting that PBC provides an additional low-variance representation-level signal that helps the detector learn robust boundaries under few-shot sampling.

\textbf{3) RAG and PBC are complementary for the best accuracy--robustness trade-off.}
RAG strengthens the attacker to provide more challenging humanization-oriented supervision, while $\mathcal{L}_{\mathrm{PBC}}$ helps the detector convert such supervision into stable representation learning; combining them yields the best joint results.
For example, on OUTFOX at 32-shot, \detectorname~achieves \textbf{95.64} accuracy and \textbf{2.36} Avg ASR, whereas \textbf{w/o RAG} reaches \textbf{92.20} accuracy and \textbf{15.66} Avg ASR, and \textbf{w/o $\mathcal{L}_{\mathrm{PBC}}$} yields \textbf{91.72} accuracy and \textbf{13.79} Avg ASR.

\subsection{Out-Of-Distribution (OOD) Experiment}
\label{sec:ood}
\begin{table*}[t]
	\centering
	\renewcommand\arraystretch{1.2}
	\footnotesize
	\resizebox{\textwidth}{!}{%
		\begin{tabular}{c c c c c c c c c}
			\toprule
			\textbf{Dataset} & \textbf{Shot} & \textbf{Metric}
			& \textbf{RoBERTa-Base} & \textbf{CoCo$^{\dagger}$} & \textbf{\pecola$^{\dagger}$} & \textbf{RADAR*} & \textbf{\modelname*} & \textbf{\detectorname~(Ours)*} \\
			\midrule
			
			\multirow{10}{*}{\textbf{OUTFOX}} 
			& \multirow{2}{*}{\textit{32}} & \textit{Acc} $\uparrow$
			& 50.90$_{\pm 3.55}$ & 50.07$_{\pm 0.06}$ & 49.58$_{\pm 0.73}$ & 35.40$_{\pm 0.00}$ & 49.32$_{\pm 8.88}$ & \textbf{87.03}$_{\bm{\pm 3.27}}$ \\
			&  & \textit{F1} $\uparrow$
			& 11.74$_{\pm 16.36}$ & 66.70$_{\pm 0.03}$ & 1.34$_{\pm 2.32}$ & 4.34$_{\pm 0.00}$ & 45.54$_{\pm 39.45}$ & \textbf{88.05}$_{\bm{\pm 2.48}}$ \\
			
			\cdashline{2-9}[1pt/1pt]
			& \multirow{2}{*}{\textit{64}} & \textit{Acc} $\uparrow$
			& 49.79$_{\pm 1.42}$ & 44.58$_{\pm 13.41}$ & 51.01$_{\pm 1.59}$ & 35.40$_{\pm 0.00}$ & 45.59$_{\pm 5.92}$ & \textbf{86.82}$_{\bm{\pm 2.21}}$ \\
			&  & \textit{F1} $\uparrow$
			& 8.92$_{\pm 8.98}$ & 43.99$_{\pm 22.72}$ & 22.97$_{\pm 38.95}$ & 4.34$_{\pm 0.00}$ & 17.27$_{\pm 28.59}$ & \textbf{87.79}$_{\bm{\pm 2.08}}$ \\
			
			\cdashline{2-9}[1pt/1pt]
			& \multirow{2}{*}{\textit{128}} & \textit{Acc} $\uparrow$
			& 60.40$_{\pm 15.12}$ & 54.41$_{\pm 7.82}$ & 57.81$_{\pm 7.97}$ & 35.40$_{\pm 0.00}$ & 55.35$_{\pm 9.28}$ & \textbf{85.45}$_{\bm{\pm 5.13}}$ \\
			&  & \textit{F1} $\uparrow$
			& 40.25$_{\pm 40.83}$ & 25.16$_{\pm 22.56}$ & 49.84$_{\pm 26.10}$ & 4.34$_{\pm 0.00}$ & 17.12$_{\pm 28.81}$ & \textbf{86.84}$_{\bm{\pm 4.00}}$ \\
			
			\cdashline{2-9}[1pt/1pt]
			& \multirow{2}{*}{\textit{256}} & \textit{Acc} $\uparrow$
			& 48.40$_{\pm 2.16}$ & 60.97$_{\pm 5.32}$ & 50.46$_{\pm 5.04}$ & 35.40$_{\pm 0.00}$ & 56.25$_{\pm 1.82}$ & \textbf{87.09}$_{\bm{\pm 1.96}}$ \\
			&  & \textit{F1} $\uparrow$
			& 12.58$_{\pm 11.00}$ & 41.00$_{\pm 16.65}$ & 51.96$_{\pm 5.54}$ & 4.34$_{\pm 0.00}$ & 34.63$_{\pm 9.62}$ & \textbf{87.53}$_{\bm{\pm 0.87}}$ \\
			
			\cdashline{2-9}[1pt/1pt]
			& \multirow{2}{*}{\textbf{\textit{Avg.}}} & \textit{Acc} $\uparrow$
			& 52.37$_{\pm 5.56}$ & 52.51$_{\pm 6.65}$ & 52.21$_{\pm 3.83}$ & 35.40$_{\pm 0.00}$ & 51.63$_{\pm 6.47}$ & \textbf{86.60}$_{\bm{\pm 3.14}}$ \\
			&  & \textit{F1} $\uparrow$
			& 18.37$_{\pm 19.29}$ & 44.21$_{\pm 15.49}$ & 31.53$_{\pm 18.23}$ & 4.34$_{\pm 0.00}$ & 28.64$_{\pm 26.62}$ & \textbf{87.55}$_{\bm{\pm 2.36}}$ \\
			
			\bottomrule
		\end{tabular}%
	}
	\caption{\textbf{OOD detection performance on HC3 dataset.} We report the mean and standard deviation of accuracy (Acc) and F1 score (F1) over 3 random seeds (mean$_{\pm \mathrm{std}}$). $^{\dagger}$ denotes few-shot-tailored detectors, and $^{*}$ indicates adversarially trained detectors. The best result in each row is \textbf{bolded}.}
	\vspace{-1.2em}
	\label{tab:ood_outfox_to_hc3_model_based}
\end{table*}

To assess the out-of-distribution (OOD) generalization of \detectorname, we train detectors on one dataset and evaluate them on a disjoint, unseen dataset.
Specifically, we train on OUTFOX and test on HC3, which is collected from different sources and contains no overlapping instances with the OUTFOX corpus, thereby probing cross-domain transfer under a realistic distribution shift.
For a fair comparison under OOD shift, we focus on model-based detectors, which can be trained under the same few-shot protocol.
HC3 evaluation uses the same label mapping and preprocessing as in-domain experiments.

Table~\ref{tab:ood_outfox_to_hc3_model_based} reports the OOD results and leads to two observations.
\textbf{1) \detectorname~achieves strong and stable OOD transfer across all shot sizes.}
\detectorname~consistently attains \textbf{86--87} accuracy and \textbf{87--88} F1 across all few-shot settings, with an overall average of \textbf{86.60} accuracy and \textbf{87.55} F1.
Notably, it remains strong even at 32-shot, reaching $\textbf{87.03}_{\bm{\pm 3.27}}$ accuracy and $\textbf{88.05}_{\bm{\pm 2.48}}$ F1, suggesting that \detectorname~does not rely on extensive in-domain supervision to transfer to an unseen distribution.
\textbf{2) Baselines degrade substantially under domain shift, and some exhibit degenerate behaviors.}
In contrast, most baselines show near-chance accuracy and/or highly unstable F1 on HC3 when trained on OUTFOX.
For example, RoBERTa-Base averages $52.37_{\pm 5.56}$ accuracy with a large-variance $18.37_{\pm 19.29}$ F1, indicating unreliable decision boundaries under domain shift.
CoCo and \pecola\ also exhibit limited transferability (Avg.\ accuracy $\approx 52$) with substantial variance across shot sizes.
Moreover, RADAR degenerates to a nearly constant behavior ($35.40_{\pm 0.00}$ accuracy and $4.34_{\pm 0.00}$ F1), suggesting that the learned decision rule fails to adapt under OOD shift.

We hypothesize that the strong OOD robustness of \detectorname~is enabled by its adversarial training mechanism.
By introducing diverse, humanization-oriented adversarial rewrites during training and regularizing representations with PBC, \detectorname~is encouraged to rely less on dataset-specific artifacts and to learn more stable cues that transfer better across domains.
This interpretation is consistent with our ablation results and training dynamics analysis in Appendix~\ref{sec:ablation} and Appendix~\ref{sec:training_dynamics}.

\subsection{TPR@FPR=1\% in Main Result}

\label{sec:tpr_fpr1_main}

While accuracy and F1 reflect overall classification quality and are already reported in the main results, evaluating \textit{TPR at a low false-positive rate} is essential for security-sensitive deployments where even a small FPR can be costly (e.g., moderation triage, plagiarism screening, or forensic auditing).
Moreover, when reporting Acc/F1, detectors often rely on tuned decision thresholds and may apply post-hoc calibration on a validation split to select an operating point.
In contrast, TPR@FPR=1\% is an operating-point metric derived from the score ranking: we sweep thresholds on the \emph{test set} and report the TPR at the threshold that attains 1\% FPR.
This evaluation is largely invariant to \emph{monotonic} score calibrations that preserve ranking.
We report TPR@FPR=1\% in Table~\ref{tab:fs_mgt_results_tpr_fpr1} and highlight three insights:
\textbf{1) Best overall low-FPR performance.}
On the overall average, \detectorname~achieves the highest \textbf{80.81} TPR@FPR=1\%.
It improves over the strongest metric-based detector Fast DetectGPT (\textbf{68.07}) by \textbf{+12.74} points, and also surpasses the best competing model-based detector \modelname\ (\textbf{76.29}) by \textbf{+4.52} points.
Across the 16 dataset--shot settings, \detectorname~obtains the best mean performance in \textbf{8/16} settings and ranks within the top-2 in \textbf{12/16}, demonstrating robust low-FPR generalization across diverse benchmarks.
\textbf{2) High recall under strict false-positive constraints.}
\detectorname~maintains high recall at stringent operating points on multiple datasets.
On DetectRL, it is near-saturated across all shots (Avg.\ \textbf{99.97}, reaching \textbf{100.00} at 64/256-shot), indicating extremely high sensitivity even when FPR is capped at 1\%.
On OUTFOX, \detectorname~already achieves \textbf{91.34} at 64-shot and scales to \textbf{95.77} at 256-shot, showing that the proposed training recipe remains effective in a practically relevant low-FPR setting.
Notably, on the challenging RAID benchmark, \detectorname~exhibits a strong scaling trend from $33.07_{\pm 27.01}$ (32-shot) to $83.33_{\pm 3.58}$ (256-shot), a gain of \textbf{+50.26} points, and becomes the best-performing method at 128/256-shot, suggesting that \detectorname~can satisfy strict low-FPR constraints once modest supervision is available.
\textbf{3) Metric-based detectors can lead in a few low-shot corners, but \detectorname~remains the strongest model-based option.}
We observe that some metric-based detectors are competitive or even superior in a small number of settings, most notably on RAID at 32/64-shot where Fast DetectGPT attains \textbf{69.96} while model-based detectors are generally weaker.
A similar pattern appears on SemEval at 32-shot where Fast DetectGPT achieves \textbf{66.37}, slightly higher than \detectorname~(\textbf{63.35}).
Despite these few-shot corner cases, \detectorname~is consistently the strongest (or among the strongest) \emph{model-based} detector under low-FPR evaluation: for instance, its RAID Avg.\ (\textbf{63.14}) outperforms \modelname\ (\textbf{56.47}) by \textbf{+6.67} points.

\begin{table*}[t]
	\centering
	\renewcommand\arraystretch{1.6}
	\footnotesize
	\resizebox{\textwidth}{!}{%
		\begin{tabular}{c c c c c c c c c c c c}
			\toprule
			\multirow{2}{*}{\textbf{Dataset}} & \multirow{2}{*}{\textbf{Shot}} & \multirow{2}{*}{\textbf{Metric}}
			& \multicolumn{3}{c}{\textbf{Metric-based Detectors}}
			& \multicolumn{6}{c}{\textbf{Model-based Detectors}} \\
			\cmidrule(lr){4-6}\cmidrule(lr){7-12}
			& & & \textbf{Fast DetectGPT} & \textbf{Binoculars} & \textbf{LRR}
			& \textbf{RoBERTa-Base} & \textbf{\pecola$^{\dagger}$} & \textbf{CoCo$^{\dagger}$} & \textbf{RADAR*} & \textbf{GREATER*} & \textbf{\detectorname~(Ours)*} \\ 
			\midrule
			\multirow{5}{*}{\textbf{DetectRL}} & \textit{32} & \textit{TPR@FPR=1\%} $\uparrow$ & 73.65$_{\pm 0.00}$ & 75.45$_{\pm 0.00}$ & 40.82$_{\pm 0.00}$ & 84.38$_{\pm 3.14}$ & 78.42$_{\pm 1.10}$ & 88.93$_{\pm 2.09}$ & \underline{97.85$_{\pm 2.49}$} & 91.02$_{\pm 8.18}$ & \textbf{99.97}$_{\bm{\pm 0.06}}$ \\
			& \textit{64} & \textit{TPR@FPR=1\%} $\uparrow$ & 73.65$_{\pm 0.00}$ & 75.45$_{\pm 0.00}$ & 40.82$_{\pm 0.00}$ & \underline{99.90$_{\pm 0.00}$} & 92.68$_{\pm 3.64}$ & 95.80$_{\pm 4.93}$ & 99.87$_{\pm 0.15}$ & 99.84$_{\pm 0.06}$ & \textbf{100.00}$_{\bm{\pm 0.00}}$ \\
			& \textit{128} & \textit{TPR@FPR=1\%} $\uparrow$ & 73.65$_{\pm 0.00}$ & 75.45$_{\pm 0.00}$ & 40.82$_{\pm 0.00}$ & 98.54$_{\pm 1.37}$ & \textbf{99.97}$_{\bm{\pm 0.06}}$ & 99.84$_{\pm 0.06}$ & 99.71$_{\pm 0.20}$ & \underline{99.93$_{\pm 0.11}$} & 99.90$_{\pm 0.17}$ \\
			& \textit{256} & \textit{TPR@FPR=1\%} $\uparrow$ & 73.65$_{\pm 0.00}$ & 75.45$_{\pm 0.00}$ & 40.82$_{\pm 0.00}$ & 99.48$_{\pm 0.73}$ & \underline{99.97$_{\pm 0.06}$} & \textbf{100.00}$_{\bm{\pm 0.00}}$ & 98.37$_{\pm 2.18}$ & \textbf{100.00}$_{\bm{\pm 0.00}}$ & \textbf{100.00}$_{\bm{\pm 0.00}}$ \\
			& \textbf{\textit{Avg.}} & \textit{TPR@FPR=1\%} $\uparrow$ & 73.65$_{\pm 0.00}$ & 75.45$_{\pm 0.00}$ & 40.82$_{\pm 0.00}$ & 95.58$_{\pm 1.31}$ & 92.76$_{\pm 1.22}$ & 96.14$_{\pm 1.77}$ & \underline{98.95$_{\pm 1.26}$} & 97.70$_{\pm 2.09}$ & \textbf{99.97}$_{\bm{\pm 0.06}}$ \\
			
			\midrule
			
			\multirow{5}{*}{\textbf{OUTFOX}} & \textit{32} & \textit{TPR@FPR=1\%} $\uparrow$ & 62.28$_{\pm 0.00}$ & 0.50$_{\pm 0.00}$ & 0.20$_{\pm 0.00}$ & \underline{73.01$_{\pm 5.23}$} & 41.70$_{\pm 29.87}$ & 72.87$_{\pm 9.05}$ & 70.63$_{\pm 13.51}$ & 64.62$_{\pm 11.03}$ & \textbf{73.57}$_{\bm{\pm 27.78}}$ \\
			& \textit{64} & \textit{TPR@FPR=1\%} $\uparrow$ & 62.28$_{\pm 0.00}$ & 0.50$_{\pm 0.00}$ & 0.20$_{\pm 0.00}$ & 75.98$_{\pm 5.60}$ & 69.50$_{\pm 6.92}$ & 86.46$_{\pm 7.15}$ & \underline{88.22$_{\pm 3.98}$} & 75.46$_{\pm 8.93}$ & \textbf{91.34}$_{\bm{\pm 1.00}}$ \\
			& \textit{128} & \textit{TPR@FPR=1\%} $\uparrow$ & 62.28$_{\pm 0.00}$ & 0.50$_{\pm 0.00}$ & 0.20$_{\pm 0.00}$ & 91.54$_{\pm 4.89}$ & 91.41$_{\pm 1.92}$ & 90.95$_{\pm 1.81}$ & 88.83$_{\pm 6.93}$ & \textbf{92.22}$_{\bm{\pm 4.00}}$ & \underline{91.75$_{\pm 0.21}$} \\
			& \textit{256} & \textit{TPR@FPR=1\%} $\uparrow$ & 62.28$_{\pm 0.00}$ & 0.50$_{\pm 0.00}$ & 0.20$_{\pm 0.00}$ & 94.37$_{\pm 4.88}$ & \textbf{96.68}$_{\bm{\pm 1.73}}$ & \underline{96.26$_{\pm 0.83}$} & 95.61$_{\pm 1.92}$ & 96.09$_{\pm 2.70}$ & 95.77$_{\pm 3.09}$ \\
			& \textbf{\textit{Avg.}} & \textit{TPR@FPR=1\%} $\uparrow$ & 62.28$_{\pm 0.00}$ & 0.50$_{\pm 0.00}$ & 0.20$_{\pm 0.00}$ & 83.73$_{\pm 5.15}$ & 74.82$_{\pm 10.11}$ & \underline{86.64$_{\pm 4.71}$} & 85.82$_{\pm 6.59}$ & 82.10$_{\pm 6.67}$ & \textbf{88.11}$_{\bm{\pm 8.02}}$ \\
			
			\midrule
			
			\multirow{5}{*}{\textbf{RAID}} & \textit{32} & \textit{TPR@FPR=1\%} $\uparrow$ & \textbf{69.96}$_{\bm{\pm 0.00}}$ & \underline{63.57$_{\pm 0.00}$} & 26.27$_{\pm 0.00}$ & 25.33$_{\pm 22.29}$ & 1.95$_{\pm 0.68}$ & 15.23$_{\pm 11.48}$ & 8.33$_{\pm 11.17}$ & 24.06$_{\pm 30.68}$ & 33.07$_{\pm 27.01}$ \\
			& \textit{64} & \textit{TPR@FPR=1\%} $\uparrow$ & \textbf{69.96}$_{\bm{\pm 0.00}}$ & 63.57$_{\pm 0.00}$ & 26.27$_{\pm 0.00}$ & 19.82$_{\pm 18.27}$ & 5.01$_{\pm 2.74}$ & 21.78$_{\pm 25.26}$ & 29.56$_{\pm 26.30}$ & 57.71$_{\pm 1.29}$ & \underline{64.32$_{\pm 13.68}$} \\
			& \textit{128} & \textit{TPR@FPR=1\%} $\uparrow$ & \underline{69.96$_{\pm 0.00}$} & 63.57$_{\pm 0.00}$ & 26.27$_{\pm 0.00}$ & 67.81$_{\pm 9.39}$ & 22.17$_{\pm 11.68}$ & 44.21$_{\pm 16.23}$ & 24.71$_{\pm 35.80}$ & 68.23$_{\pm 4.80}$ & \textbf{71.84}$_{\bm{\pm 2.29}}$ \\
			& \textit{256} & \textit{TPR@FPR=1\%} $\uparrow$ & 69.96$_{\pm 0.00}$ & 63.57$_{\pm 0.00}$ & 26.27$_{\pm 0.00}$ & \underline{79.56$_{\pm 6.99}$} & 53.26$_{\pm 9.33}$ & 72.49$_{\pm 1.04}$ & 34.05$_{\pm 35.06}$ & 75.88$_{\pm 4.82}$ & \textbf{83.33}$_{\bm{\pm 3.58}}$ \\
			& \textbf{\textit{Avg.}} & \textit{TPR@FPR=1\%} $\uparrow$ & \textbf{69.96}$_{\bm{\pm 0.00}}$ & \underline{63.57$_{\pm 0.00}$} & 26.27$_{\pm 0.00}$ & 48.13$_{\pm 14.24}$ & 20.60$_{\pm 6.11}$ & 38.43$_{\pm 13.50}$ & 24.16$_{\pm 27.08}$ & 56.47$_{\pm 10.40}$ & 63.14$_{\pm 11.64}$ \\
			
			\midrule
			
			\multirow{5}{*}{\textbf{SemEval}} & \textit{32} & \textit{TPR@FPR=1\%} $\uparrow$ & \textbf{66.37}$_{\bm{\pm 0.00}}$ & \underline{63.77$_{\pm 0.00}$} & 13.17$_{\pm 0.00}$ & 60.38$_{\pm 6.35}$ & 12.34$_{\pm 6.68}$ & 13.37$_{\pm 0.88}$ & 58.23$_{\pm 4.74}$ & 39.81$_{\pm 29.59}$ & 63.35$_{\pm 11.30}$ \\
			& \textit{64} & \textit{TPR@FPR=1\%} $\uparrow$ & 66.37$_{\pm 0.00}$ & 63.77$_{\pm 0.00}$ & 13.17$_{\pm 0.00}$ & 57.62$_{\pm 14.69}$ & 46.03$_{\pm 11.50}$ & 71.52$_{\pm 7.52}$ & 72.80$_{\pm 5.84}$ & \underline{73.00$_{\pm 4.48}$} & \textbf{73.24}$_{\bm{\pm 7.19}}$ \\
			& \textit{128} & \textit{TPR@FPR=1\%} $\uparrow$ & 66.37$_{\pm 0.00}$ & 63.77$_{\pm 0.00}$ & 13.17$_{\pm 0.00}$ & 70.92$_{\pm 6.98}$ & 69.62$_{\pm 4.82}$ & 72.33$_{\pm 6.14}$ & 56.84$_{\pm 32.63}$ & \textbf{78.96}$_{\bm{\pm 7.07}}$ & \underline{72.62$_{\pm 15.72}$} \\
			& \textit{256} & \textit{TPR@FPR=1\%} $\uparrow$ & 66.37$_{\pm 0.00}$ & 63.77$_{\pm 0.00}$ & 13.17$_{\pm 0.00}$ & 70.80$_{\pm 19.40}$ & 70.43$_{\pm 2.44}$ & 72.87$_{\pm 2.17}$ & 76.76$_{\pm 4.80}$ & \textbf{83.79}$_{\bm{\pm 7.41}}$ & \underline{78.84$_{\pm 18.30}$} \\
			& \textbf{\textit{Avg.}} & \textit{TPR@FPR=1\%} $\uparrow$ & 66.37$_{\pm 0.00}$ & 63.77$_{\pm 0.00}$ & 13.17$_{\pm 0.00}$ & 64.93$_{\pm 11.86}$ & 49.61$_{\pm 6.36}$ & 57.52$_{\pm 4.18}$ & 66.16$_{\pm 12.00}$ & \underline{68.89$_{\pm 12.14}$} & \textbf{72.01}$_{\bm{\pm 13.13}}$ \\
			
			\midrule
			
			\textbf{Overall} & \textbf{\textit{Avg.}} & \textit{TPR@FPR=1\%} $\uparrow$ & 68.07$_{\pm 0.00}$ & 50.82$_{\pm 0.00}$ & 20.12$_{\pm 0.00}$ & 73.09$_{\pm 8.14}$ & 59.45$_{\pm 5.95}$ & 69.68$_{\pm 6.04}$ & 68.77$_{\pm 11.73}$ & \underline{76.29$_{\pm 7.82}$} & \textbf{80.81}$_{\bm{\pm 8.21}}$ \\
			\bottomrule
		\end{tabular}%
	}
	\caption{\textbf{TPR@FPR=1\% across 4 datasets and shot sizes.} We report the mean and standard deviation of TPR@FPR=1\% over 3 random seeds for each experiment (mean$_{\pm \mathrm{std}}$). $^{\dagger}$ denotes few-shot-tailored detectors, and $^{*}$ indicates adversarially trained detectors. The best result(s) in each row are \textbf{bolded}, and the second-best result(s) are \underline{underlined}.}
	\label{tab:fs_mgt_results_tpr_fpr1}
\end{table*}

\subsection{Human Evaluation of Adversarial Example Generation Strategy}
\label{sec:human}
\begin{table*}[t]
	\centering
	\renewcommand\arraystretch{1.25}
	\footnotesize
	\resizebox{1\textwidth}{!}{%
		\begin{tabular}{c c c c c c}
			\toprule
			\textbf{Generation Strategy} & \textbf{H1} & \textbf{H2} & \textbf{H3} & \textbf{Mean} & \textbf{Significance} \\
			\midrule
			Direct Generate & $2.660\pm1.327$ & $3.330\pm1.356$ & $2.390\pm1.294$ & $2.793\pm1.326$ & $2.13\mathrm{e}{-12}^{*} \,/\, \text{--}$ \\
			\cdashline{1-6}[1pt/1pt]
			Direct Prompt   & $2.660\pm1.208$ & $3.600\pm1.279$ & $2.700\pm1.345$ & $2.987\pm1.277$ & $4.05\mathrm{e}{-09}^{*} \,/\, 1.16\mathrm{e}{-01}$ \\
			Few-shot Prompt & $2.720\pm1.207$ & $3.320\pm1.230$ & $2.510\pm1.337$ & $2.850\pm1.258$ & $9.16\mathrm{e}{-12}^{*} \,/\, 5.74\mathrm{e}{-01}$ \\
			RAG (Ours)      & $3.560\pm1.149$ & $4.120\pm0.856$ & $3.010\pm1.283$ & $3.563\pm1.096$ & $1.18\mathrm{e}{-01} \,/\, 1.16\mathrm{e}{-09}^{*}$ \\
			\cdashline{1-6}[1pt/1pt]
			HWT             & $3.700\pm1.068$ & $4.110\pm0.777$ & $3.370\pm1.405$ & $3.727\pm1.083$ & $\text{--} \,/\, 2.13\mathrm{e}{-12}^{*}$ \\
			\bottomrule
		\end{tabular}%
	}
	\caption{\textbf{Human evaluation of adversarial example generation strategies (human-likeness, 1--5).}
		H1--H3 are three independent raters scoring how human-like each text is (higher is more human-like).
		\textbf{Mean} reports the average of the three rater means and the average of the three rater standard deviations.
		\textbf{Significance} reports two paired Wilcoxon signed-rank tests (two-sided) in the format \emph{vs HWT / vs Direct Generate}.
		${}^{*}$ indicates $p<0.01$ (significant difference); ``--'' denotes non-applicable comparisons (the row itself is the reference).}
	\label{tab:human_eval_generation_strategy}
	\vspace{-1.2em}
\end{table*}

\noindent \textbf{Experiment Settings.}
We conduct a human study to assess whether our RAG-based generation can truly produce more human-like adversarial texts.
We randomly sample 100 human-written texts (HWTs) and generate corresponding texts with four strategies:
(1) \textbf{Direct Generate}, where we directly use an LLM to generate the rewrite version of a given HWT without any additional prompting;
(2) \textbf{Direct Prompt}, where we use a single instruction prompt to encourage human-like writing;
(3) \textbf{Few-shot Prompt}, where we provide one MGT/HWT pair as in-context guidance;
and (4) \textbf{RAG (Ours)}, where we use retrieval-augmented generation with curated context.
Three annotators (senior undergraduate students; English as a second language) independently rate each text on a 1--5 human-likeness scale (higher indicates more human-like).
All texts (generated and HWT) are fully shuffled and presented in random order; annotators receive a short training session and are shown three example pairs before annotation.
Each annotator finishes within two hours and is compensated at twice the local minimum hourly wage.
For significance testing, we adopt the following protocol: for each item $i$, we first aggregate the three ratings as $\bar{y}_i(G_k)=\frac{1}{3}\sum_{r\in\{\mathrm{H1,H2,H3}\}} y_{i,r}(G_k)$, and then apply a two-sided paired Wilcoxon signed-rank test between $\{\bar{y}_i(G_k)\}_{i=1}^{100}$ and the reference set (HWT or Direct Generate), marking $p<0.01$ as significant.

\noindent \textbf{Experiment Results.}
Table~\ref{tab:human_eval_generation_strategy} shows that \textbf{RAG (Ours)} is perceived as substantially more human-like than the other generation strategies.
RAG achieves the highest mean score (\textbf{3.563}), closing most of the gap to real HWT (\textbf{3.727}).
More importantly, RAG is \emph{not} significantly different from HWT under our strict criterion ($p=1.18\mathrm{e}{-01}\ge 0.01$), while it is significantly different from Direct Generate ($p=1.16\mathrm{e}{-09}<0.01$).
This indicates that human raters consider RAG outputs markedly more human-like than naive LLM generations, and statistically indistinguishable from human writing at $p<0.01$.
In contrast, \textbf{Direct Prompt} and \textbf{Few-shot Prompt} provide limited improvement.
Both remain significantly different from HWT ($p=4.05\mathrm{e}{-09}$ and $p=9.16\mathrm{e}{-12}$), yet neither differs significantly from Direct Generate ($p=1.16\mathrm{e}{-01}$ and $p=5.74\mathrm{e}{-01}$).
Together, these results suggest that retrieval-guided context is a key factor for generating stronger, more human-like adversarial examples, rather than relying on lightweight prompting alone.

\begin{table*}[!htbp]
	\centering
	\renewcommand\arraystretch{1.4}
	\resizebox{1\textwidth}{!}{%
		\begin{tabular}{c m{0.7\textwidth} >{\centering\arraybackslash}m{0.05\textwidth} >{\centering\arraybackslash}m{0.05\textwidth} >{\centering\arraybackslash}m{0.05\textwidth} >{\centering\arraybackslash}m{0.06\textwidth}}
			\hline
			\textbf{Method} & \textbf{Text} & \textbf{H1} & \textbf{H2} & \textbf{H3} & \textbf{Mean} \\
			\hline
			
			\textbf{Raw MGT} &
			Genshin Impact is an open-world action RPG featuring elemental combat, gacha characters, and an expansive fantasy world called Teyvat. &
			2 & 3 & 1 & 2.00 \\
			\hline
			
			\textbf{Modify} &
			Genshin Impact is an open-world action RPG featuring combat, gacha and lore, set in an expansive fantasy world called Teyvat. &
			1 & 2 & 1 & 1.33 \\
			\hline
			
			\textbf{Paraphrasing} &
			THE GENSHIN IMPACT, an open world action RPG, involves elemental combat, Gacha-style characters and a vast fantasy land called Teyvat. &
			2 & 3 & 2 & 2.33 \\
			\hline
			
			\textbf{Back-translation} &
			Genshin Impact is an open-world action RPG with elementary combat, Gacha characters and an expansive fantasy world called Teyvat. &
			1 & 3 & 1 & 1.67 \\
			\hline
			
			\textbf{Direct Prompt} &
			An immersive experience, Genshin Impact is an open-world action role-playing game. It presents players with a dynamic elemental combat system,
			a character acquisition model based on the gacha mechanic, and a vast fantasy setting known as Teyvat for exploration. &
			2 & 3 & 2 & 2.33 \\
			\hline
			
			\textbf{Few-shot Prompt} &
			Genshin Impact is an open-world action RPG featuring elemental combat, gacha characters, and an expansive fantasy world called Teyvat.
			Here, players can explore a rich, dynamic world filled with a variety of mystical creatures, mystical beasts, and powerful magical artifacts,
			all of which contribute to the diverse gameplay of the title. &
			3 & 4 & 2 & 3.00 \\
			\hline
			
			\textbf{RAG (Ours)} &
			You know, there's this game I keep coming back to---Genshin Impact. It drops you into this breathtaking, wide-open fantasy realm called Teyvat,
			where you're free to wander and discover. The real magic, for me, is in the combat. You get to mix and match elemental powers in these dazzling,
			almost musical combos. And the characters? Sure, you collect them through the gacha system, but they all have such rich stories. It feels less like
			checking off features and more like stepping into a living, breathing world that’s just waiting to be explored. &
			4 & 5 & 3 & 4.00 \\
			\hline
			
		\end{tabular}%
	}
	\caption{\textbf{Case study of attacks on the samples in \figref{fig:fig1}.}
		The scoring rubric and annotation protocol are the same as \secref{sec:human}.
		\textbf{Mean} is computed as the average of \textbf{H1--H3}.}
	\vspace{-0.4cm}
	\label{tab:case_study_results}
\end{table*}

\subsection{Case Study}
\label{sec:case}
\noindent \textbf{Experiment Settings.}
To qualitatively understand how different attacks and generation strategies change the perceived human-likeness of machine-generated text, we present a representative example in Table~\ref{tab:case_study_results} (the same sample as in \figref{fig:fig1}).
We compare 3 perturbation-based attacks (\textbf{Modify}, \textbf{Paraphrasing}, \textbf{Back-translation}) used in \secref{sec:attackresult} against prompting-based generation strategies (\textbf{Direct Prompt}, \textbf{Few-shot Prompt}) and our \textbf{RAG (Ours)} generator.
All outputs are rated by the same three annotators and follow the identical scoring rubric and annotation protocol described in \secref{sec:human}, where each text is scored on a 1--5 scale indicating how human-like it appears.

\noindent \textbf{Experiment Results.}
Table~\ref{tab:case_study_results} illustrates a clear gap between \emph{surface-level} perturbations and \emph{humanization-oriented} generation.
Starting from the raw MGT baseline (Mean $=2.00$), perturbation-based attacks either fail to improve human-likeness or even make the text more suspicious.
For example, \textbf{Modify} slightly reshuffles local phrasing but remains terse and list-like, receiving the lowest score (Mean $=1.33$).
Similarly, \textbf{Back-translation} introduces unnatural word choice (e.g., ``elementary combat'') and preserves the original template-like structure, yielding only a marginal Mean of $1.67$.
\textbf{Paraphrasing} changes casing and wording but still reads like a feature enumeration, resulting in a limited improvement (Mean $=2.33$), comparable to the raw text.

Prompting-based strategies provide moderate gains but remain constrained by the same ``feature-summary'' writing style.
\textbf{Direct Prompt} expands the description with more formal phrasing, yet the output still resembles a polished product synopsis and receives Mean $=2.33$.
\textbf{Few-shot Prompt} further increases length and injects additional content (e.g., creatures and artifacts), which improves human-likeness to Mean $=3.00$, but parts of the added details feel generic and less grounded, limiting the perceived authenticity.

In contrast, \textbf{RAG (Ours)} yields the most human-like output (Mean $=4.00$).
The text not only introduces richer content but also exhibits distinctly human cues: (i) a personal stance (``there's this game I keep coming back to''), (ii) narrative flow with discourse markers (``You know,'' ``for me''), (iii) subjective impressions and metaphors (``almost musical combos,'' ``living, breathing world''), and (iv) balanced treatment of potentially machine-like signals such as ``gacha'' by embedding them into a coherent personal experience rather than listing them as features.
These characteristics collectively make the generation read less like a templated description and more like natural human commentary, explaining the consistently higher ratings across annotators.

Overall, this case study supports the main takeaway of our human evaluation: naive perturbations and lightweight prompting tend to preserve the underlying ``LLM-summary'' signature, whereas retrieval-guided generation can substantially shift the writing style toward human-like discourse, producing stronger and more realistic adversarial examples for training robust detectors.

\subsection{Analysis of Training Dynamics}
\label{sec:training_dynamics}

\begin{figure*}[htbp]
	\centering
	\resizebox{1\textwidth}{!}{%
		\includegraphics{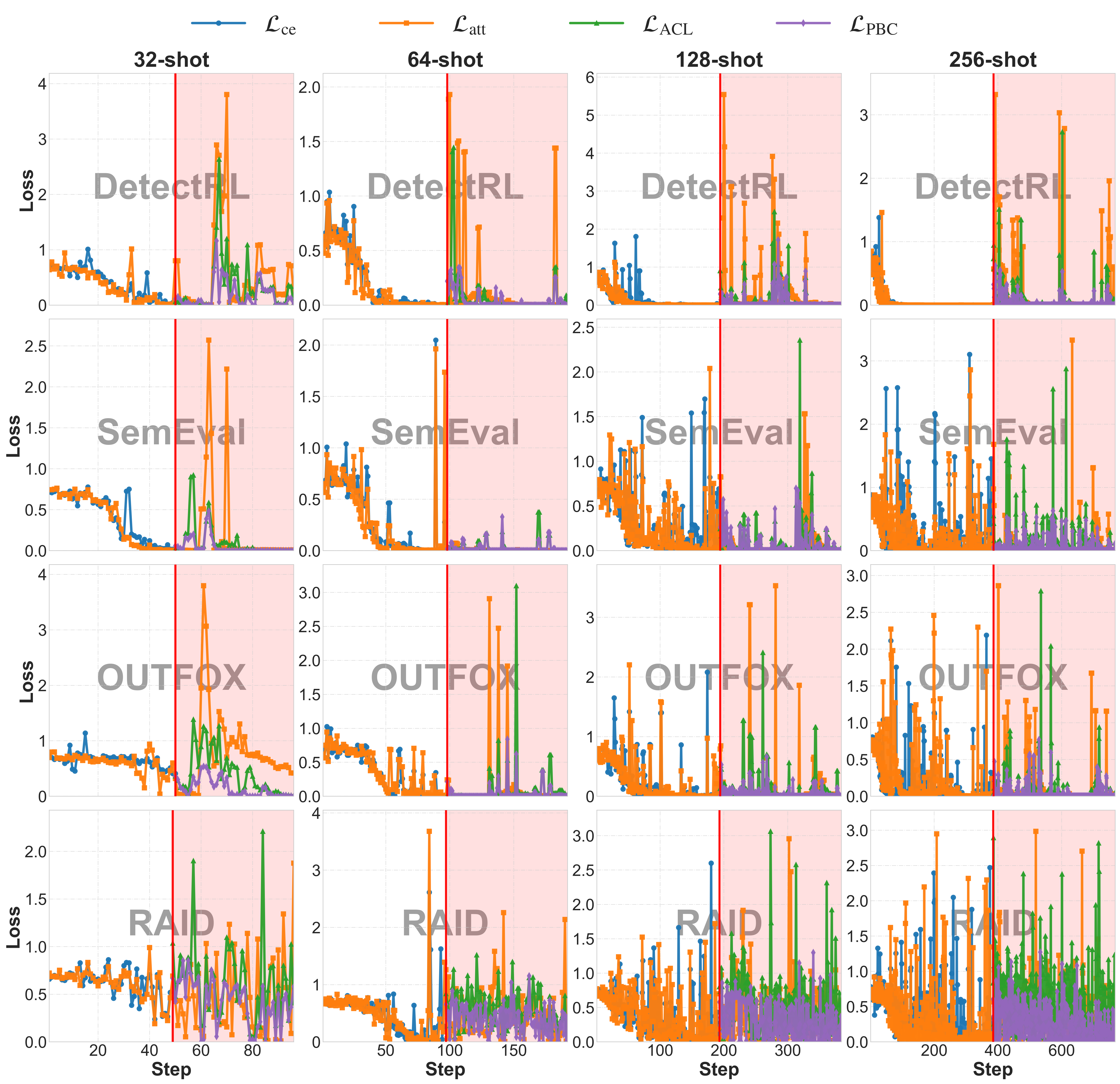}
	}
	\caption{\textbf{Loss curves over training steps.}
		We report step-wise losses for both the detector and the attacker, including the detector cross-entropy loss during pre-training ($\mathcal{L}_{\text{ce}}$), the attacker loss during pre-training ($\mathcal{L}_{\text{att}}$), the detector ACL loss during adversarial training ($\mathcal{L}_{\text{ACL}}$), the detector PBC loss during adversarial training ($\mathcal{L}_{\text{PBC}}$), and the attacker loss during adversarial training ($\mathcal{L}_{\text{att}}$).
		The red vertical line marks the stage transition, and the adversarial training stage is highlighted with a light-red background.
	}
	\label{fig:loss}
\end{figure*}

To better understand the convergence and stability of our adversarial training procedure---and to examine how the surrogate attacker evolves throughout training---we visualize the step-wise loss curves across \emph{all datasets} and \emph{all shot sizes} in Figure~\ref{fig:loss}.
From Figure~\ref{fig:loss}, we summarize three key observations.

\noindent \textbf{(i) Stable optimization across the stage transition.}
A common concern in few-shot adversarial training is that switching from clean supervised learning to adversarial updates can destabilize optimization, especially when data are scarce.
Across all datasets and shot sizes, our training remains well-behaved: losses stay within a controlled range, and the stage transition does not induce catastrophic spikes, divergence, or long-lasting oscillations.
Even in the 32-shot setting, where stochasticity is typically amplified, the detector adapts quickly after entering the adversarial stage and maintains a stable loss profile thereafter.
Overall, these curves suggest that our training schedule and objectives avoid abrupt distribution shifts that could derail learning.

\noindent \textbf{(ii) Alternating updates yield a meaningful attacker--detector interaction.}
Within the adversarial stage, we observe a consistent coupling between the attacker-side and detector-side objectives.
In many settings, fluctuations in the attacker loss $\mathcal{L}_{\text{att}}$ coincide with corresponding changes in the detector’s adversarial loss $\mathcal{L}_{\text{ACL}}$.
This pattern is expected under our alternating-update design: as the attacker produces rewrites that are harder for the current surrogate (and thus harder to imitate the target detector’s label-only feedback), the detector is exposed to more challenging humanized samples and places greater emphasis on maintaining correct decisions under humanization, which is reflected by $\mathcal{L}_{\text{ACL}}$.
Subsequently, as the detector adapts, both losses settle into a more stable regime.
Importantly, the sustained dynamics across datasets and shot sizes indicate that the surrogate attacker does not trivially collapse to a fixed, non-informative rewriting pattern; instead, the interaction resembles an adaptive curriculum in which the detector is continually trained on progressively more challenging humanized candidates.

\noindent \textbf{(iii) PBC provides a fast, low-variance signal that complements $\mathcal{L}_{\text{ACL}}$.}
Finally, Figure~\ref{fig:loss} shows clear convergence behavior within each stage.
During pre-training, $\mathcal{L}_{\text{ce}}$ decreases smoothly, indicating effective fitting on clean supervision.
After switching to adversarial training, $\mathcal{L}_{\text{ACL}}$ and $\mathcal{L}_{\text{PBC}}$ rapidly move toward a stable regime rather than remaining persistently high in low-data regimes.
Notably, $\mathcal{L}_{\text{PBC}}$ typically becomes active early and stabilizes quickly.
This behavior aligns with our motivation for PBC: as a pairwise, margin-based geometric constraint, it provides a low-variance learning signal that does not rely on large batches or many in-batch negatives, which are often unreliable in few-shot settings.
Empirically, the early stabilization of $\mathcal{L}_{\text{PBC}}$ is accompanied by faster stabilization of $\mathcal{L}_{\text{ACL}}$, suggesting that PBC helps regularize representation geometry and facilitates more efficient convergence of the main adversarial classification objective.

Overall, Figure~\ref{fig:loss} supports that our framework yields (1) stable optimization across the pre-training--to--adversarial transition, (2) an effective attacker--detector interaction under alternating updates with label-only feedback, and (3) reliable convergence with PBC providing an additional, well-conditioned signal that complements ACL in few-shot adversarial training.






\end{document}